\documentclass[conference, compsoc]{IEEEtran}

\usepackage{graphicx} \usepackage{url}
\usepackage{amsmath}
\usepackage{amsfonts, amssymb}
\usepackage{xcolor}
\usepackage{amsthm}
\usepackage{comment}
\usepackage{booktabs}
\usepackage{siunitx}
\usepackage{algorithm, algorithmic}
\usepackage{multirow}
\usepackage{colortbl}
\usepackage{xcolor} \mathchardef\mhyphen="2D

\def\ba{{\mathbf{a}}}
\def\bb{{\mathbf{b}}}

\newcommand{\enc}[1]{\textsf{Enc}\left(#1\right)}
\newcommand{\ct}{\textsf{ct}}

\newcommand{\sk}{\textsf{sk}}

\newboolean{publishedversion}
\setboolean{publishedversion}{false}

\def\ZZ{{\mathbb{Z}}}
\def\RR{{\mathbb{R}}}
\def\CC{\mathbb{C}}
\def\R{{\mathcal{R}}}

\def\Ecd{\textsf{Ecd}}

\def\Add{\textsf{Add}}
\def\PCMult{\textsf{PCMult}}

\def\Rot{\textsf{Rot}}

\def\PCMult{\textsf{PCMult}}

\def\sk{\mathsf{sk}}
\def\rk{\mathsf{rk}}

\def\rlk{\mathsf{rlk}}
\def\ksk{\mathsf{ksk}}

\def\ct{\mathsf{ct}}

\def\iDFT{\textrm{iDFT}}

\newcommand{\FFLASBibkeyhack}[7]{}
\newcommand{\HEaaNBibkeyhack}[5]{}

\NewDocumentCommand{\opnorm}{sO{}m}{\IfBooleanTF{#1}{$\left|\opnormkern\left|\opnormkern\left|
    #3
    \right|\opnormkern\right|\opnormkern\right|
  $}{
    \mathopen{#2|\opnormkern #2|\opnormkern #2|}
    #3
    \mathclose{#2|\opnormkern #2|\opnormkern #2|}
  }}
\newcommand{\opnormkern}{\mkern-1.5mu\relax}

\def\bb{{\bf b}}
\def\ba{{\bf a}}

\def\bm{{\bf m}}
\def\pt{{\mathsf{pt}}}
\def\ct{{\mathsf{ct}}}
\def\ctres{{\mathsf{ct}_{\mathrm{res}}}}
\def\Rot{{\mathsf{Rot}}}
\def\RR{{\mathbb{R}}}

\def\etok{{\mathsf{etok}}}
\def\ntok{{\mathsf{ntok}}}
\def\ptok{{\mathsf{ptok}}}
\def\PCMM{{\mathrm{PCMM}}}
\def\PCMM{\text{PCMM}}
\def\CCMM{{\mathrm{CCMM}}}
\def\CC{{\mathbb{C}}}
\def\RR{{\mathbb{R}}}
\def\iDFT{{\mathrm{iDFT}}}
\def\ecdslot{{\mathsf{Ecd}_{\mathrm{slot}}}}
\def\ecdcoeff{{\mathsf{Ecd}_{\mathrm{coeff}}}}

\def\dcd{{\mathsf{Dcd}}}
\def\enc{{\mathsf{Enc}}}
\def\dec{{\mathsf{Dec}}}
\def\Mult{{\mathsf{Mult}}}
\def\RS{{\mathsf{Rescale}}}
\def\dnum{{\mathsf{dnum}}}
\def\KS{{\mathsf{KeySwitch}}}
\def\v{{\mathsf{v}}}
\def\m{{\mathsf{m}}}

\newtheorem{lemma}{Lemma}
\newtheorem{theorem}{Theorem}

\newcommand{\SinC}{\mathsf{SinC}}

\newcommand{\rotrow}{\mathsf{rot}_{R}}
\newcommand{\rotcol}{\mathsf{rot}_{C}}
\definecolor{myblue}{RGB}{10,30,210}

\definecolor{draft-color}{RGB}{40,100,60}

\definecolor{draft-color-2nd}{RGB}{90,70,130}

 \usepackage[shortlabels]{enumitem}
\usepackage[T1]{fontenc}
\usepackage{graphicx}
\usepackage{libertine}
\usepackage{multirow}
\usepackage{tikz}
\usepackage{pgfplots}
\usepackage{stmaryrd} 

\title{Scaling up FHE-based Privacy-Preserving ML: \\ Higher Throughput, Longer Inputs for LLama-3-8B}

\author{
\IEEEauthorblockN{
Jaiyoung Park\IEEEauthorrefmark{1}, 
Sejin Park\IEEEauthorrefmark{2}\textsuperscript{,}\IEEEauthorrefmark{3}, 
Jai Hyun Park\IEEEauthorrefmark{4}, 
Jung Ho Ahn\IEEEauthorrefmark{1}, 
Jung Hee Cheon\IEEEauthorrefmark{2}\textsuperscript{,}\IEEEauthorrefmark{3}, \\
Guillaume Hanrot\IEEEauthorrefmark{4}, 
Jung Woo Kim\IEEEauthorrefmark{2}, 
Minje Park\IEEEauthorrefmark{2}, and 
Damien Stehlé\IEEEauthorrefmark{4}
}

\IEEEauthorblockA{\IEEEauthorrefmark{1}\textit{Graduate School of Convergence Science and Technology, Seoul National University}, Seoul, South Korea}
\IEEEauthorblockA{\IEEEauthorrefmark{2}\textit{CryptoLab, Inc.}, Seoul, South Korea}
\IEEEauthorblockA{\IEEEauthorrefmark{3}\textit{Department of Mathematics, Seoul National University}, Seoul, South Korea}
\IEEEauthorblockA{\IEEEauthorrefmark{4}\textit{CryptoLab, Inc.}, Lyon, France}
}

\date{June 2026}

\begin{document}

\maketitle

\begin{abstract}
As large language models (LLMs) become ubiquitous, privacy concerns pertaining to inference inputs keep growing. In this context, fully homomorphic encryption (FHE) has emerged as a primary cryptographic solution to provide non-interactive confidential LLM inference. Existing solutions scale poorly with the input token length, and hence focus either on small models or larger models with a small number of input tokens. They also suffer from the existence of large outlier values. These values have a strong impact on the evaluation of non-linear layers, leading to large-degree polynomial approximation and thus heavy evaluation costs. We scale up FHE-based LLM inference in two different directions. 

First, we accelerate the FHE-based LLM inference in the setting of 128 input tokens. To do so, we adopt and revisit ML-techniques (token prepending and orthogonal rotations) to mitigate the impact of the outliers on FHE evaluation of non-linear layers. Separately, we devise a new polynomial evaluation method for sparsely-packed ciphertexts, which we leverage to speed up our homomorphic SoftMax implementation. We combine these ingredients with several recent fast homomorphic linear algebra techniques for the linear layers, achieving improved efficiency for 128 encrypted input tokens. 

Second, we expand the prompt size of FHE-based private LLM inference up to thousands of input tokens when the context is benign and only part of the input is sensitive. Our framework then assumes that only the last part of the prompt is encrypted. Processing such a prompt requires handling a plaintext-plaintext component (identical to the fully clear case) and a ciphertext-ciphertext component (identical to the fully encrypted case). In addition, it requires handling a wide homomorphic computation corresponding to a plaintext-ciphertext component specific to our proposed framework. To address this component, we devise a dedicated homomorphic linear algebra algorithm, which is a main building block for a tailored shallow homomorphic attention circuit designed to minimize the bootstrapping costs. 

Based on these ingredients, we describe a CKKS-based end-to-end implementation of \mbox{Llama-3-8B} private inference. On a cluster of 8~NVIDIA \mbox{RTX PRO 6000} GPUs, inference for an input of 128 encrypted tokens takes 20s for summarization and 18s (per output token) for generation. As a comparison, the state-of-the-art solution by Jayashankar \emph{et al.} [arxiv'25] performs the same summarization task in 295s on 8~much costlier H100 GPUs. In the case of a heterogeneous public/private input of 4096 tokens, the last 128 being encrypted, our implementation takes 64s for summarization and 22s (per output token) for generation on 8~NVIDIA \mbox{RTX PRO 6000} GPUs.   
\end{abstract}

\section{Introduction}
When querying a Large Language Model (LLM), clients typically send their queries to a remote server that to run the model inference. Because these queries and their generated results often contain sensitive data, significant privacy concerns naturally emerge. To 
address this, recent research has actively pursued privacy-preserving LLM inference, primarily relying on Secure Multi-Party Computation (MPC), Fully Homomorphic Encryption (FHE), or a combination of both.

MPC-based solutions~\cite{HLL+23,WFZ+24,KC25,CXS+25,DLZW+25,GJM+24,ZYHC+25,LHG+25,XLL+25,YZL23} suffer from two major drawbacks: they require many rounds of communication with the client during the computation, and they rely on the strict assumption that computing parties do not collude. In contrast, FHE-based computation  incurs minimal interaction, but its efficient deployment requires taming a potentially high computational cost. In this work, we focus on inference using the CKKS FHE scheme~\cite{CKKS17}.

Prior works remain limited in both model size and performance. Several studies~\cite{CBH+22,MYJK25,PLL25,YCD+25,ZWSH+25} consider BERT, and only a few address the considerably larger Llama models~\cite{Llama}. When they do so, it is typically in simulated settings and for only a few input tokens (e.g., 8~in~\cite{ZYHC+25,ZWSH+25}).

Even for these restricted workloads, obtaining an implementation with practical performance remains challenging and requires high-end GPUs. For example, the NEXUS protocol from~\cite{ZYHC+25} requires an amortized 51.84\,s per token, corresponding to 414.72\,s for the initial prompt processing phase (the prefill stage) of \mbox{Llama-3-8B} using four A100 GPUs. The MOAI protocol from~\cite{ZWSH+25} requires 224.53s to process an ~8~token prefill of \mbox{Llama-3-8B} on a single A100 GPU. 
More recently, a systems work~\cite{JKSZS25} demonstrated encrypted inference for \mbox{Llama-3-8B} with~128 input tokens, taking 295s using~8~H100 GPUs and 134s using 8~B200~GPUs. 

Given these runtimes, increasing the number of input tokens seems challenging: some components have a cost that grows linearly in the number of input tokens, while others (at the core of the attention phase) have costs that grow quadratically. Furthermore, increasing the model size and the number of tokens leads to robustness difficulties. Larger models lead to larger ranges of values for intermediate variables~\cite{sckl24}, 
because of outliers that largely exceed the average magnitudes. These directly impact the efficiency of FHE inference, through increased degrees of polynomial approximations, increased ciphertext modulus consumption, and ciphertext maintenance  costs.

\subsection{Contributions}
To address these challenges, we develop a collection of algorithmic and system techniques to integrate and optimize up-to-date algorithmic solutions for each encrypted LLM evaluation task, targetting practical FHE-based LLM inference.
We realize these techniques in \emph{Sylph}, an end-to-end implementation of FHE-based privacy-preserving Transformer inference. Our main contributions are as follows:

\begin{itemize}
\item \textbf{FHE-aware model calibration and outlier mitigation.}
We identify activation outliers as a major source of inefficiency in encrypted LLM inference and develop model-level calibration techniques that reduce the magnitude of outliers without retraining. This approach reduces the dynamic range required during homomorphic evaluation, improving both efficiency and robustness for modern Transformer models.

\item \textbf{Heterogeneous public/private prompt processing for long-context inference.}
We introduce an inference framework that jointly processes a heterogeneous prompt consisting of a public and a private segment within a single Transformer execution. This design enables efficient long-context inference while preserving strong confidentiality guarantees for sensitive user inputs.

\item\textbf{FHE circuit design with various ciphertext formats.}
We design optimized FHE circuits that incorporate model calibration for both homogeneous and heterogeneous prompts. We efficiently integrate diverse state-of-the-art FHE algorithms, each relying on different ciphertext formats. To satisfy the format requirements without introducing extra overhead, we systematically fuse the ciphertext format conversions into existing FHE operations such as bootstrapping. Furthermore, we propose new FHE algorithms for linear algebra and SoftMax evaluation. 

\item \textbf{Scalable distributed execution for encrypted Transformers.}
We design a distributed runtime for homomorphic Transformer inference that scales across multiple GPUs. Through careful coordination of computation, communication, and ciphertext placement, Sylph achieves strong scaling efficiency up to eight GPUs, enabling practical deployment of modern LLMs and long-context inference.
\end{itemize}

Table~\ref{tab:cc-quick-result} summarizes the performance of Sylph and prior HE inference systems under a unified 128-token private-input setting. Combining
the techniques described above, Sylph completes prefill in 20s on eight RTX Pro 6000 GPUs and 26s on four B200 GPUs, substantially outperforming prior systems.
Sylph also completes decoding in 18s on eight RTX Pro 6000 GPUs. For heterogeneous prompts with 3,968 public tokens and 128 encrypted tokens, Sylph finishes prefill in 64s and decoding in 22s. These results significantly narrow the gap between encrypted inference and practical LLM deployment.

\begin{table}[h]
\centering
\small
\setlength{\tabcolsep}{4pt}
\renewcommand{\arraystretch}{1.1}
\caption{
Timings (in seconds) for HE inference systems under various hardware configurations.
Cerium timings are extracted from~\cite{JKSZS25}.
$^\dagger$ MOAI results are estimates (the workload exceeds the memory capacity of GPUs).
Input settings per benchmark are noted in the first column. RP6000 refers to RTX PRO 6000. N/A means that the corresponding hardware was not accessible.
}
\label{tab:cc-quick-result}
\begin{tabular}{llccccc}
\toprule
Benchmark & Phase & GPU & 1$\times$ & 2$\times$ & 4$\times$ & 8$\times$ \\
\midrule
\multirow{4}{*}{\shortstack[l]{Sylph (Ours)\\{\scriptsize 128 priv.}}}
& Decode  & B200      & 24  & 17  & 12  & N/A  \\
& Decode  & RP6000  & 31  & 24  & 19  & 18 \\
& Prefill & B200      & 87  & 47  & 26  & N/A \\
& Prefill & RP6000 & 116 & 63  & 36  & 20 \\
\midrule
\multirow{4}{*}{\shortstack[l]{Sylph (Ours)\\{\scriptsize 3968 pub.}\\{\scriptsize +128 priv.}}}
& Decode  & B200      &  - & -  & 15  &  N/A  \\
& Decode  & RP6000      & - & -  & 23  &  22  \\
& Prefill & B200      & -  & -  &  90   &  N/A  \\
& Prefill & RP6000  & -  & -  &  129   &  64  \\
\midrule
\multirow{2}{*}{\shortstack[l]{Cerium\\{\scriptsize 128 priv.}}}
& Prefill & B200  & 253 & 215 & 152 & 134 \\
& Prefill & H100  & 698 & 580 & 346 & 295 \\
\midrule
\multirow{2}{*}{\shortstack[l]{MOAI$^\dagger$\\{\scriptsize 128 priv.}}}
& Prefill & B200     & 5,963    & - & - & - \\
& Prefill & RP6000 & 22,275  & - & - & - \\
\bottomrule
\end{tabular}
\end{table}

\subsection{Technical overview}
FHE-based LLM inference generates encrypted output tokens from the given encrypted input tokens. Scaling it up creates two main difficulties: the increased magnitudes of outliers in the scaled models must be taken care of to provide robust FHE inference, and the ciphertext-ciphertext computations can incur a computational blow-up.

\medskip\noindent\textbf{Tailoring LLM models.} 
Large language models are known to exhibit activation outliers, which become a major efficiency bottleneck under FHE. Outliers increase the cost of FHE computation in two ways. First, FHE bootstrapping must support higher relative precision: even though some dedicated algorithms exist~\cite{BCCKKS22,CKSS25}, their cost is significantly higher than regular CKKS bootstrapping. Second, as CKKS supports only addition and multiplication, computing non-arithmetic functions is performed via polynomial approximation; larger outliers enlarge the approximation domain, leading to higher-degree polynomials and additional homomorphic operations~\cite{CKP22}.

To mitigate this challenge, we incorporate model-level techniques that suppress activation outliers before homomorphic execution. First, following~\cite{xtchl24}, we prepend a small set of ``faulty tokens’’ to the input, producing an initial Key-Value cache (a representation of past tokens used to speed up generation) that reduces extreme activations during inference. Second, following~\cite{quarot24}, we apply adaptive orthogonal rotations that distribute the influence of outlier coordinates across many dimensions, reducing their peak magnitude. 
Based on these calibration and characterization results, we design efficient low-degree polynomial approximations and evaluation techniques for nonlinear operators, further reducing the cost of Transformer evaluation.

\medskip\noindent\textbf{Slim polynomial evaluation for SoftMax}. 
We optimize the homomorphic SoftMax algorithm from~\cite{CHKPS24} by exploiting the sharp control of the ranges of inputs and intermediate values in the calls to SoftMax in the \mbox{Llama} models. Recall that the algorithm from~\cite{CHKPS24} consists of two tracks: a wide main track and a slim-but-deep auxiliary track. 

We optimize the auxiliary track by leveraging
the fact that the corresponding ciphertexts are sparsely packed, i.e., their message dimension is much smaller than the ciphertext capacity. 
For this purpose, we introduce a slim polynomial evaluation algorithm for sparsely packed ciphertexts. The main idea is to use the superfluous message slots for the purpose of the computation. This is achieved by (recursively) decomposing the polynomial to be evaluated as a sum of squares of two other polynomials of half degree, and evaluating these two polynomials in parallel.

\medskip\noindent\textbf{Heterogeneous prefill stages}
A major obstacle to scale FHE inference to long-context is the cost of prefill, the initial step that handles the prompt. The cost of prefill scales quadratically  with the input length because attention evaluates pairwise interactions between input token. Consequently, processing thousands of encrypted prompt tokens becomes prohibitively expensive.

To address this challenge, Sylph adopts a heterogeneous public/private prompt strategy that separates public and private prompt segments. This design targets applications where a long public context is followed by a short sensitive user query.

The public part of the prompt is processed in cleartext while  only the sensitive suffix is processed under encryption. Because of the split, a novel operation appears during the attention stage, where the encrypted tokens are processed against the cleartext tokens. We call this operation \emph{PC-attention}.
When dealing with a large public prompt, the cost of this new PC-attention dominates the total prefill execution. Optimizing this specific stage becomes critical for overall efficiency.

\medskip\noindent\textbf{Bootstrapping-frugal PC-attention}.  
Long public contexts lead to a highly unbalanced workload. PC-attention becomes a massively wide operation (up to 128 ciphertexts in our experiments). In contrast, the remaining parts of the encrypted computation scale only with the sensitive prompt and remain moderate in size. Handling this asymmetry efficiently requires avoiding bootstrapping operations during the wide PC-attention phase. 

To achieve this, we design PC-attention to be extremely shallow in terms of homomorphic multiplications. We introduce a new plaintext-ciphertext matrix multiplication (PCMM) algorithm adapted from~\cite{JKLS18}. Furthermore, by maintaining precise control over variable ranges, we implemented a shallow and lightweight SoftMax evaluation adapted from~\cite{CHKPS24}. As a result, PC-attention can be executed with only a small number of bootstrapping operations.

\medskip\noindent\textbf{New PCMM for PC-attention}. 
The state-of-the-art PCMM algorithm of~\cite{BCHPS24} is shallow and efficient, but operates under coefficient encoding and is thus tightly coupled to bootstrapping. The PCMM algorithms of~\cite{JKLS18} and~\cite{MYJK25} are more flexible, but require~$2$ or~$3$ multiplicative levels. The PCMM algorithm in~\cite{PLL25} is shallow, flexible (it operates on slot encoding), and computationally cheaper than~\cite{MYJK25,JKLS18}. 
However, except for~\cite{BCHPS24}, all the above algorithms (including~\cite{PLL25}) require manipulation of the plaintext matrices. Previous algorithms choose to precompute and store the transformed matrices to save computation time. When performing PCMMs at a large ciphertext modulus, storing many copies of the plaintext matrix however becomes prohibitive.

In order to handle the high-modulus PCMM task, we devise a new PCMM algorithm specifically for the PC-attention context. Building on~\cite{JKLS18, MYJK25}, it retains the qualities of~\cite{PLL25} while having low memory footprint. In contrast to~\cite{JKLS18}, we allow our PCMM algorithm to flexibly update the layout structure of the underlying FHE ciphertexts, enabling seamless integration with the attention operation. We adopt Baby-step Giant-step (BSGS) to achieve the same asymptotic complexity as~\cite{PLL25}. The proposed PCMM algorithm stores a single copy of the plaintext matrix without and processes it on the fly during computation, making the memory footprint  reasonable despite the large modulus.

\medskip\noindent\textbf{FHE circuit design with various ciphertext formats}.
To put it all together, we design optimized FHE circuits for both short homogeneous private prompts and long heterogeneous public/private prompts. 
We integrate state-of-the-art homomorphic linear algebra algorithms (\cite{BCHPS24,CKL25,rhombus,HS14}) with the proposed algorithms for SoftMax evaluation and PCMM in PC-attention. 

The main challenge for the design is the various ciphertext formats for each FHE algorithm. The ciphertext formats comprise ring degree, encoding structures, and packing layouts. 
For example, PCMM algorithms from~\cite{BCHPS24} use coefficient-encoding and various FHE ring degrees. The CCMM algorithm from~\cite{CKL25} exploits SinC-encoding, while most other algorithms are slot-encoding based. Also, the proposed PCMM algorithm uses different packing layouts, which are specialized for PC-attention. 
The naive solution to address this issue is to insert format conversion layers between every incompatible algorithm, but they incur a substantial cost overhead. 

Instead, we systematically fuse most format conversions into existing FHE operations. We show that, for our parameters, SinC encoding~\cite{CKL25} is compatible with ring-switching techniques, thereby supporting various ring degrees. Subsequently, we fuse most encoding conversions into bootstrapping, as suggested in~\cite{BCHPS24,Park25}. 
Furthermore, we design a packing layout minimizing the required conversions. This allows us to execute the optimized algorithms within a single computation flow while retaining their operator-level efficiency.

\medskip\noindent\textbf{Scalable distributed execution.}
The preceding techniques reduce the computational cost of encrypted Transformer inference, but practical deployment still requires an  efficient execution and compact memory footprint across multiple GPUs. To this end, Sylph co-designs data partitioning and communication placement to minimize both memory footprint and communication overhead. 

The first challenge is that ciphertexts vary significantly in size throughout the execution pipeline, particularly before and after bootstrapping. Our design carefully places communication at stages where ciphertexts reside at the lowest modulus level and avoids redundant bootstrapping to achieve both communication and computation efficiency.
The second challenge is that efficient operators often come at the cost of increased memory footprint. Sylph carefully balances these tradeoffs while preserving compact ciphertext and weight representations. 

Overall, these design principles enable strong scaling across multiple GPUs while maintaining a compact memory footprint. As a result, Sylph can serve modern LLMs within the memory capacity of easily accessible GPU systems (RTX family).

\section{Preliminaries}
We use boldface letters for vectors and capital letters for matrices. The vector spaces~${\mathbb R}^n$ and ${\mathbb C}^n$ are equipped as rings with coordinate-wise addition and multiplication (denoted by~$\odot$).
We let $\lfloor \cdot \rceil$ denote rounding to nearest (with an arbitrary rule for ties).  The base-2 logarithm is denoted by~$\log$. 
Throughout the paper, we abbreviate the multiplication between a plaintext matrix and a ciphertext matrix as $\PCMM$, that between ciphertext matrices as $\CCMM$, that between a plaintext vector and a ciphertext matrix as PCMv, and that between ciphertext vector and matrix as CCMv.

\subsection{The CKKS scheme}
CKKS~\cite{CKKS17} is an RLWE-based FHE scheme implementing arithmetic over approximate  (real or complex) numbers. Its main parameter is the ring degree $N = 2^n$ (typically~$n = 16$). It has message space $\CC^{N/2}$, 
while its plaintext space is $\R_N = \ZZ[X]/(X^N+1)$ -- we shall simply use $\R$ when there is no ambiguity on the ring degree.

\subsubsection{Encodings}\label{se:ecd}
Let $\zeta = \exp(\pi\sqrt{-1}/N)$ be a primitive $(2N)$-th root of unity. We define the map $\textrm{iDFT}_N: \CC^{N/2} \rightarrow \RR[X]/(X^N+1)$ by
\[
  P = \iDFT_N((\bm)_{0\le i < N/2}) \Leftrightarrow P(\zeta^{5^i}) = m_i, \ 0\le i < N/2 \enspace.
\]
CKKS maps the message space to the plaintext space using one of the following two encoding functions, parameterized by a large integer $\Delta$ called the \emph{scaling factor} and which drives the numerical precision. On input $\bm = (m_i)_{0\le i < N/2}$, we define
\begin{align*}
\ecdslot(\bm) & = \lfloor \Delta \cdot \iDFT(\bm) \rceil\enspace,\\  
\ecdcoeff(\bm) & =\sum_{i=0}^{N/2-1} \left(\lfloor \Delta \textrm{Re}(m_i) \rceil + X^{N/2} \lfloor \Delta \textrm{Im}(m_i) \rceil\right) X^i\enspace. 
\end{align*}

Besides slot and coefficient encoding, we also use an intermediate encoding called \emph{Slots-in-Coefficients} (SinC), introduced in~\cite{CKL25}. 

Let $k$ be a divisor of $N$, and let us identify $\R_k = \mathbb{Z}[Y]/(Y^k+1)$ with a subring of $\R_N$ by mapping $Y$ to $X^{N/k}$. The $\textrm{iDFT}_k$ map, with root of unity $\xi = \zeta^{N/k}$ sends $\mathbb{C}^{k/2}$ to $\R_k$. Given $\mathbf{z}\in \mathbb{C}^{N/2}$, we define $\mathbf{z}^{(i)} = (z_{ik/2}, \dots, z_{(i+1)k/2-1}) \in \mathbb{C}^{k/2}$ and 
\begin{equation}\label{eq:sinc_ecd}
\Ecd_{\SinC_{k,N}}(z) = \sum_{i=0}^{N/k-1} \lfloor \Delta \cdot \textrm{iDFT}_k(\mathbf{z}^{(i)}) \rceil X^i \in \R_N\enspace.
\end{equation}

Conversion between encodings can be performed by homomorphically evaluating the relevant DFT or iDFT maps; we discuss these transforms in Section~\ref{sse:bts}. 

When the message to be encoded has fewer coordinates than~$N/2$, we call it \emph{sparsely-packed} or, for short, \emph{slim}. This terminology is extended to plaintexts (and ciphertexts) if their underlying messages are slim. Algebraically, if~$N'$ is the smallest power of two that is no smaller than the number of coordinates, we have the plaintext live in a subring~$\R_{N'}$ of~$\R_N$; equivalently, when using slot-encoding the underlying message is periodic, of the form
\[
\underbrace{(x_0, \dots, x_{N'/2-1}, x_0, \dots, x_{N'/2-1}\dots, x_0, \dots, x_{N'/2-1})}_{N/N' \textrm{ times}}\enspace.
\]

In this work, as we have no use for complex numbers, we shall use the conjugate-invariant version of CKKS~\cite{KS18}; the latter can be seen as a variant on top of ordinary CKKS, giving access to $N$ real slots rather than $N/2$ complex slots while keeping the fact that slot encoding is an approximate ring homomorphism. This induces a number of technicalities for which the interested reader is referred to~\cite{KS18}. 

We review CKKS functionalities (encryption, decryption, operations) in Appendix~\ref{app:CKKS}.

\subsubsection{Bootstrapping}\label{sse:bts}
CKKS ciphertexts are elements of $\R_Q^2$, where $Q$ is a modulus which decreases whenever the ciphertext undergoes a multiplication. 
At some point, a ciphertext can no longer be multiplied unless it undergoes modulus restoration, called bootstrapping. CKKS bootstrapping~\cite{CHKKS18} is a complex and costly operation, which often accounts for a large proportion of the total cost of deep homomorphic circuits. As such, organizing the computation so as to minimize the number of bootstrappings is often a key to the design of efficient homomorphic algorithms. 

Parts of the bootstrapping algorithm manipulate the internal encoding of a ciphertext, by homomorphically evaluating the DFT and iDFT maps. By interrupting temporarily the bootstrapping flow or adapting it (in the case of the $\SinC$ encoding), one  gains access to the coefficient encoding or $\SinC$ encoding and use their algebraic properties to design efficient algorithms.

\subsection{Homomorphic Linear Algebra}
\label{se:hom_linalg}
The evaluation of linear algebra operations over encrypted data has been the subject of substantial research, with spectacular progress in recent years. We review the solutions that we use in four main categories: PCMM, CCMM, PCMv and CCMv.

\begin{itemize}
    \item \textbf{PCMM from BLAS}: In~\cite{BCHPS24}, the authors propose fast PCMM algorithms by leveraging reductions from PCMM to clear matrix multiplications, inspired by verifiable PCMM in~\cite{LZ22}. By offloading most computations to fast BLAS libraries, they achieve significant speedups. Furthermore, their algorithms consume the multiplicative depth optimally. Consequently, the algorithms can be performed with small FHE moduli, and thus, they are highly advantageous in minimizing the memory footprint. 

    \item \textbf{CCMM from BLAS}: Extending this highly efficient paradigm, \cite{Park25} proposes fast CCMM algorithms for large matrices, enabling CCMM from fast BLAS libraries. Subsequently, \cite{CKL25} further extends it to smaller matrices when many CCMMs are batched. The algorithms in ~\cite{CKL25} utilize the SinC encoding, which is an intermediate encoding between coefficient and slot encodings, which can thus be accessed by fusing the CCMM algorithm with CKKS bootstrapping. Note that the algorithms in~\cite{BCHPS24} and~\cite{Park25} utilize coefficient encoding, which is also easily accessible during bootstrapping. 

    \item \textbf{PCMM/CCMM with slot encoding}: When it is not feasible to fuse PCMM/CCMM with bootstrapping, slot encoding is more beneficial. The notable work~\cite{JKLS18} proposes CCMM algorithms for slot-encoded ciphertexts with a minimized number of CKKS key-switchings. Most existing PCMM and CCMM algorithms with slot encoding are based on the algorithm in~\cite{JKLS18}. 

    \item \textbf{PCMv and CCMv}: For PCMv, a seminal work~\cite{HS14} provides an optimized solution. A critical optimization is the application of the Baby-step Giant-step (BSGS) technique, which significantly reduces the number of key-switching operations during PCMv. The algorithm is designed for slot encoding, while a recent work \cite{rhombus} achieves a similar result for coefficient encoding. 
    The algorithm in~\cite{HS14} can be adapted to CCMv, but the BSGS technique cannot be directly applied, and it incurs a higher number of key-switchings than PCMv. 
\end{itemize}

\subsection{Homomorphic SoftMax Evaluation}
\label{se:shallow_Softmax}
We briefly review the homomorphic SoftMax algorithm of~\cite{CHKPS24}, which serves as the basis of our implementation. Let $(x_i)_{0\le i < d}$ be the input to SoftMax.

\begin{itemize}
    \item Based on input range estimates, the input coordinates are translated to $[-M, 0]$  and scaled to $x'_i \in [-M/2^k, 0]$ for some real number~$M$ and integer~$k$ (for all~$0 \leq i <d$); 

    \item Using polynomial approximation over $[-M/2^k, 0]$, one computes $y_i^{(0)} = \exp(x'_i)$ for $0\le i < d$; 

    \item One computes $y_{i}^{(j)} = (y_i^{(j-1)} / 
    \|(y_\ell^{(j-1)})_{0\le \ell < d} \|)^2$ for $1\le j \le k$; the inverse square root caused by the Euclidean norm in the denominator needs not be very accurate except at the last iteration~(see~\cite{CHKPS24}); 

    \item One returns $(y_i^{(k)})_{0 \le i < d}$.
\end{itemize}

The algorithm consists of a wide main path that evaluates exponential and repeated squaring operations, combined with a narrow auxiliary path that computes the normalization factors. The inverse square roots in the auxiliary track are evaluated using polynomial approximations~\cite{trefethen2019atap}, and can be refined by additional Newton iterations when higher accuracy is required.

Since the auxiliary track operates on only $d$ values, whereas the main track operates on all SoftMax coordinates, polynomial evaluation is often performed on \emph{sparsely-packed ciphertexts} (see Section~\ref{se:ecd}).

\subsection{Llama Architecture and Inference}

Llama is a decoder-only Transformer architecture With RMSNorm, RoPE (rotary positional-embedding), multi-head attention, and a SwiGLU feed-forward network. We use Llama-3~\cite{dubey2024llama} as a representative modern LLM, as its architecture is shared by many recent open-weight LLMs.

Llama generates text autoregressively, meaning it produces one new token per forward pass. The initial pass, referred to as the \emph{prefill} stage, processes the entire user prompt at once to  generate the first output token. Subsequent passes are called \emph{decode} stages. 
Because each new token must "attend" to the context of all preceding tokens, a naive decoding approach would repeatedly recompute the mathematical representations (specifically, the Key (K) and Value (V) states) of the entire sequence. Modern LLMs avoid this redundancy through KV caching, which stores previously computed Key-Value states and reuses them during subsequent decoding steps.

Because of this design, the prefill and decode stages exhibit fundamentally different computational profiles. The decode stage processes a single token at a time and is dominated by matrix-vector operations. In contrast, the prefill stage processes the entire prompt at once and relies on heavier matrix-matrix operations. Moreover, attention computation during prefill scales quadratically with the input context length, making long prompts particularly expensive. Consequently, prefill dominates the overall inference cost and becomes the primary challenge for long-context FHE inference.

\subsection{Outlier Phenomenon in LLMs}
LLMs exhibit unusually large activation values during inference, a behavior known as the \emph{outlier phenomenon}~\cite{sckl24}. Such outliers are particularly problematic for FHE inference because they enlarge the approximation range of nonlinear functions, leading to higher-degree polynomial approximations.

Prior work identifies two major sources of outliers. The first is the \emph{attention sink} phenomenon~\cite{xtchl24}, where a small number of tokens attract disproportionately large attention scores, causing spikes in the model's internal values (activations). The second is the presence \emph{dimension-wise outliers}, where specific features ("hidden dimensions") consistently exhibit significantly larger values than others~\cite{quarot24,smoothquant23}.

To mitigate these effects, several activation-smoothing techniques have been proposed for LLMs. Attention-sink mitigation suppresses rare activation spikes by prepending sink tokens~\cite{xtchl24} to the prompt. For dimension-wise outliers, orthogonal 
rotation techniques redistribute the large values  across all hidden dimensions~\cite{quarot24}. Because the rotation and its inverse can be pre-computed and fused into the model's existing linear layers, this smoothing step incurs no additional cost during inference. 
Together, they reduce activation magnitudes and improve numerical stability.
We provide the visualization of Llama inference pipeline in Figure~\ref{fig:pt-alg} in Appendix~\ref{app:figure}.  \section{Accelerating Encrypted Llama Inference}

\label{se:outlier}
In this section, we present several techniques and new FHE algorithms for LLM inference on fully encrypted input tokens. By systematically integrating  effective LLM techniques and optimal, state-of-the-art FHE components, our method significantly improves the performance of FHE-based LLM inference. Specifically, we combine
prepending sink tokens~\cite{xtchl24}, 
the LLM orthogonal rotation approach~\cite{quarot24}, 
and recent homomorphic linear algebra algorithms~\cite{BCHPS24,CKL25,rhombus} with our new homomorphic SoftMax evaluation algorithms (Section~\ref{se:slim-poly}), achieving a substantial improvement over prior work for the same task.

\subsection{Model Calibration}\label{sec:model_calibration}
\subsubsection{Outlier Mitigation}
To improve FHE inference efficiency and accuracy, we perform offline outlier mitigation and noise-aware analysis.

First, we mitigate attention-sink behavior using a short sink-inducing prefix. A small set of tokens that consistently trigger attention-sink outliers is identified offline, and the resulting Key-Value states are precomputed and injected as a static KV cache during inference. Since the prefix is independent of user inputs, it can be shared across all queries without privacy concerns.

Second, we apply orthogonal rotations to mitigate dimension-wise outliers. The rotations are fused into adjacent linear projections during model preparation, restricting them to the down-proj, o-proj, and v-proj layers. As orthogonal transformations preserve model functionality, they incur no runtime overhead while redistributing activation energy more evenly across hidden dimensions.

Together, the outlier mitigation techniques reduce the numerical range of non-linear computations from approximately 2400 to below 40 (See Tables~\ref{prefix_reduction} and~\ref{prefix_rotation_ablation}).
\begin{table}[t]
\centering
\caption{Maximal input magnitudes of non-linear layers with and without prefix optimization.}
\label{prefix_reduction}
\small
\setlength{\tabcolsep}{6pt}
\begin{tabular}{lccc}
\toprule
\textbf{Position} & \textbf{Baseline} & \textbf{With Prefixing} & \textbf{Reduction} \\
\midrule
SoftMax& 39.24  & 32.78  & $-16\%$ \\
SiLU & 23.00  & 10.82  & $-53\%$ \\
RMSNorm & 2243.97  & 7.65  & $-99.7\%$ \\
\bottomrule
\end{tabular}
\end{table}

\begin{table}[t]
\centering
\caption{Maximum magnitude of the output of various projection steps after calibration (experimental estimation).}
\label{prefix_rotation_ablation}
\small
\setlength{\tabcolsep}{6pt}
\begin{tabular}{lcc}
\toprule
\textbf{Position} & \textbf{Baseline}& \textbf{Calibrated (Sec~\ref{sec:model_calibration})} \\
\midrule
down-proj & 310.56 & \textbf{1.92} \\
o-proj    & 10.12  & \textbf{1.08} \\
v-proj    & 5.89   & \textbf{4.74} \\
\bottomrule
\end{tabular}
\end{table}

\subsubsection{Precision requirements}
After applying these transformations, we estimate the required precision. For a given precision, we run the forward pass with injection of modeled CKKS noise at each operation, and compute the perplexity score~\cite{jurafsky2024speech}. We observe that by using 12 bits of precision, we match the perplexity of plaintext FP16 evaluation (see~Table~\ref{tab:precision_perplexity} in Appendix~\ref{app:perp}).

\subsubsection{Impact on circuit design}
This calibration of the numerical aspects of the model directly impacts the circuit design. The sharp control of the required accuracy allows us to set the prime sizes for multiplicative level in CKKS parameters precisely, optimizing the modulus usage. The control of the range of the input values at each stage allows us to optimize the design of non-linear layer and to avoid using the expensive high-precision bootstrapping. 

\emph{Non-linear layers}.
The resulting activation statistics and accuracy requirements are used to determine both the approximation range and target precision of nonlinear operators.  Reducing the operating range substantially lowers both polynomial degree  and execution cost. For SiLU, we decrease the polynomial degree from 83 to 31 (gaining 2 levels) and the computation time by $1.55\times$.

\emph{Bootstrapping precision}.
\label{se:extbts}
CKKS bootstrapping assumes inputs in $[-1,1]$, and its precision degrades rapidly outside this range. Although prior works~\cite{BCCKKS22,CKSS25} enable higher-precision bootstrapping over larger input ranges, they incur substantially higher evaluation costs. Instead, Sylph scales ciphertexts before bootstrapping. Specifically, given a $p$-bit bootstrapping procedure and an input bound $B$, we scale a ciphertext $\ct$ by $1/B$ so that all slot values lie within $[-1,1]$. This scaling reduces the effective precision by $\log_2 B$ bits, yielding $p-\log_2 B$ bits of precision after bootstrapping. For Llama-3-8B, we use a 20-bit bootstrapping procedure with $B=128$, resulting in 13-bit effective precision while satisfying the required input range.

\subsection{Homomorphic Execution Pipeline}
\label{se:pipeline}

Building on the model calibration techniques described in the previous section, we describe the overall computation flow of our implementation, Sylph, for fully encrypted input tokens. 
We homomorphically evaluate the prefill stage and subsequent decode stages in sequnce. The prefill stage consists of four operations as follows. 
\begin{enumerate}
    \item \textbf{QKV-projection}: pre-attention RMSNorm, QKV, and RoPE steps. 
    \item \textbf{Attention operation}: QK, SoftMax, and ScoreV steps.
    \item \textbf{O-projection}: O-projection, residual addition, and pre-FFN RMSNorm steps.
    \item \textbf{FFN operation}: SwiGLU and down-projection steps. 
\end{enumerate}
We refer to Figure~\ref{fig:ct-alg} for the visualization. Only the dark gray boxes correspond to the computational flow for the setting of this Section, where the input contains only encrypted tokens. 

For the FHE computation, QKV, O-projection, and down-projection steps correspond to PCMMs.
The RoPE step is a position-wise plaintext-ciphertext multiplication,
and QK and ScoreV steps are CCMMs. SoftMax and RMSNorm steps homomorphically evaluate the SoftMax and an inverse square-root function. We implement SwiGLU using a PCMM and a homomorphic SiLU evaluation. 

For PCMM and CCMM steps, while most existing approaches to FHE-based LLM inference exploit variants of the algorithm in~\cite{JKLS18}, we utilize BLAS-based methods~\cite{BCHPS24,Park25,CKL25,BCHPS25} outlined in Section~\ref{se:hom_linalg}. These methods consume only a single multiplicative level, and more importantly, fully leverage the high efficiency of the GPU linear algebra libraries such as ~\cite{cublas}, significantly improving homomorphic linear algebra steps. For SoftMax, we further develop the slim polynomial evaluation technique described in Section~\ref{se:slim-poly}.

The decoding stage follows the same overall structure, with matrix-matrix multiplications replaced by matrix-vector multiplications. For PCMv and CCMv, we adopt the algorithms from~\cite{rhombus} and~\cite{HS14}, respectively.

\medskip
\noindent\textbf{How to orchestrate the components?} 
Instantiating this computation flow using the CKKS scheme introduces distinct technical challenges due to the incompatible ciphertext formats across different FHE algorithms.
For the ease of discussion, we represent each intermediate ciphertext format by three attributes: (i) \emph{ring degree}, which determines the underlying polynomial ring of each ciphertext, (ii) \emph{encoding structure}, which determines how the messages are embedded into the ciphertext, e.g., as slots, coefficients, or SinC coordinates, and (iii) \emph{packing layout}, which specifies the mapping from tensor coordinates into the encoding structure. 
Table~\ref{tab:he-operators} summarizes the ciphertext formats during each operator in Sylph. Note that Batch PCMM operator is not used in this section, but it is required in Section~\ref{se:large-input}. 

\begin{table}[h]
\centering
\caption{Homomorphic linear algebra algorithms used in Sylph. Row-split denotes a layout where each matrix row is packed into one or more ciphertexts. Row and Diagonal denote row-by-row and diagonal-by-diagonal layouts, respectively.}
\label{tab:he-operators}
\resizebox{\columnwidth}{!}{
\begin{tabular}{lcccc}
\toprule
 Operator & Method & Ring Degree & Encoding & Packing Layout \\
\midrule

PCMM
    &\cite{BCHPS24}
    & $256$
    & Coeff
    & Row-split \\

PCMV
    &\cite{rhombus}
    & $4096$
    & Coeff
    & Row \\

Batch PCMM
    & Section~\ref{se:pcmm}
    & $65536$
    & Slot
    & Diagonal \\

\midrule

Batch CCMM
    & \cite{CKL25}
    & $4096$
    & SinC
    & Row \\

Batch CCMV
    & \cite{HS14}
    & $65536$
    & Slot
    & Diagonal \\

\midrule

Non-linear Ops
    & Section~\ref{se:slim-poly}
    & $65536$
    & Slot
    & -\\

\bottomrule
\end{tabular}
}
\footnotesize
\end{table}

Orchestrating all components with incompatible ciphertext formats requires  appropriate format conversion methods. To connect different ring degrees, we utilize the ring-switching technique~\cite{GHPS12} mostly with coefficient encoding as suggested in~\cite{BCHPS24}, for the PCMM/PCMv steps. In Section~\ref{sec:generalized-ring-conversion}, we extend the ring switching technique to SinC encoding, for the Batch-CCMM steps.

Regarding encoding conversions, we use standard methods, SlotToCoeff, CoeffToSlot, and their variants. Following the strategy from~\cite{BCHPS24}, we fuse those encoding conversions into bootstrapping. This fusion reduces the conversion overhead and allows heavy linear algebra to be computed at the smallest possible modulus, reducing both computation cost and memory footprint. 

\medskip
\noindent\textbf{Packing layout strategy.} 
Finally, we complete the pipeline by designing a packing layout that accomodates subsequent transformations (bit-reversal) and decompositions. This design yields the exact ciphertext structures required by the various linear algebra algorithms that we use, effectively \emph{removing the need for most conversions throughout the entire process}. The only remaining layout conversions are transpose and $\tau^2$ (discussed in Section~\ref{se:pcmm}), which we implement using algorithms in~\cite{HS14}, with BSGS optimizations. 

We focus primarily on the attention operation, which is the most complex in terms of data organization. Throughout this operation, each token is described by a vector of real values, which we refer to as \emph{channels}. At the onset of the attention phase, the total number of channels is $2^{12}$. 

During attention, these channels undergo three different projections (key, query and value); each resulting set of values is split into $n_{\textrm{heads}}$ blocks of 128 channels.  

The corresponding projected data for the 128 input tokens can be structured as a 3-dimensional tensor $(t_{i,j,k})\in \mathbb{R}^{n_{\textrm{heads}}\times 128\times 128}$, where:
\begin{itemize}
    \item $i$ is the "head" index ($0 \le i < n_{\textrm{heads}}$),
    \item $j$ is the channel index within a head ($0 \le j < 128$), and
    \item $k$ is the token index ($0 \le k < 128$).
\end{itemize}

Each such tensor is processed during the attention operation as $n_{\textrm{heads}}$ distinct matrices of dimension $128\times 128$. Our layout maps the values $t_{i, j, k}$ into the ciphertext slots following the lexicographic order of the triples $(j, i, k)$. Specifically, when enumerating slots: 
\begin{itemize}
    \item \textbf{Token index $k$ varies fastest:} A given ciphertext stores the values of one specific channel and head across a block of 128 consecutive slots.
    \item \textbf{Head index $i$ varies next:} These 128-slot blocks corresponding to a given channel across all $n_{\textrm{heads}}$ heads are packed into $128\cdot n_{\textrm{heads}}$ consecutive slots.
\end{itemize}

This data is stored into $n_{\textrm{heads}}/4$ ciphertexts of ring degree $2^{16}$. For $0\le \ell < n_{\textrm{heads}}/4$ and $0\le m < 2^{16}$, slot $s$ of the $\ell$-th ciphertext $\ct_\ell$ is defined as:
\[
\ct_{\ell}[s] = t_{\lfloor s / 2^7 \rfloor \bmod n_{\textrm{heads}},~ \ell\cdot 2^9/n_{\textrm{heads}} + \lfloor s / (2^7 n_{\textrm{heads}}) \rfloor ,~ s \bmod 2^7}.
\]

\begin{figure}
    \centering
    \includegraphics[width=0.9\columnwidth]{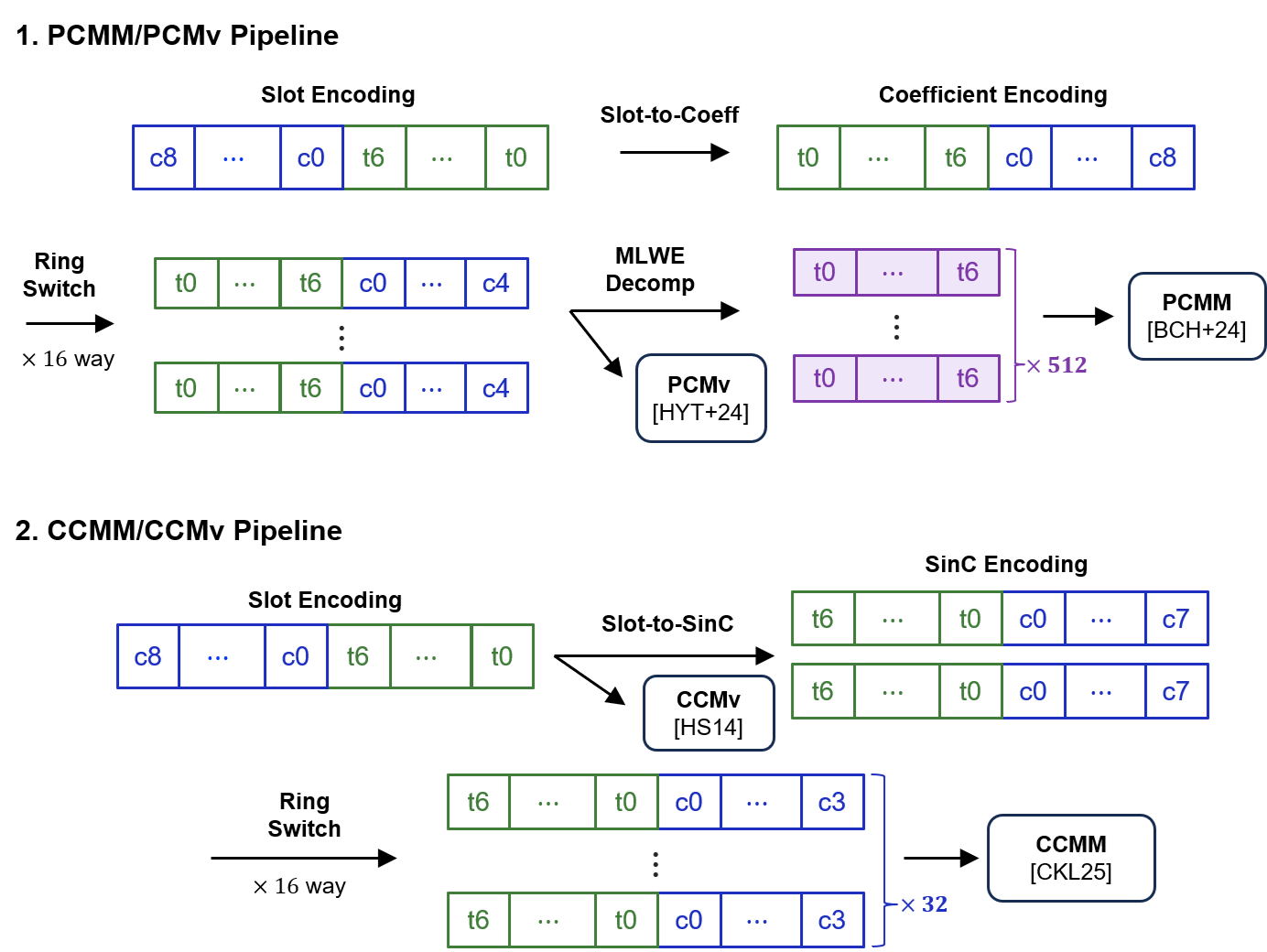}
    \caption{Ciphertext formats, packing layouts, and conversion paths used for Llama-3 evaluation. The labels $t_0,\ldots,t_6$ denote bit positions of the token coordinate for a 128-token block, not individual tokens. The labels $c_i$ correspond to (part of) the channel number. 
}
    \label{fig:pipeline}
\end{figure}

This layout is designed to accomodate the bit-reversal operation that occurs when moving to coefficient encoding, and the partial bit-reversal operation which occurs when moving to $\SinC$ encoding. Following this bit reversal operation, the roles of dimensions $i$ and $k$ are swapped (see~Fig.~\ref{fig:pipeline}).

In the case of PCMM, ring-switching followed by an MLWE decomposition keeps all the token values of one given channel co-located in one ciphertext. In the case of batch-CCMM, ring-switching after conversion to $\SinC$ encoding~(see Section~\ref{sec:generalized-ring-conversion}) ensures the same property. Simultaneously, it brings different rows of a $128\times 128$ matrix corresponding to a given head at the same positions of different ciphertexts.  
The plaintext matrices (weight matrices) and masks (RoPE) are encoded accordingly. The remaining linear algebra steps (within the FFN stage) are handled similarly, except that the pair of tensor dimensions $(i, j)$ becomes a single dimension. 
Finally, non-linear steps are only mildly affected by layout choices (when gathering terms to compute inverse square roots in SoftMax or RMSNorm, or propagating the results).

\subsection{Ring-switching in SinC encoding}
\label{sec:generalized-ring-conversion}
Unlike PCMM and PCMv, whose input ciphertext formats are low-degree coefficient encoding, CCMM from~\cite{CKL25} requires SinC encoding input, with low ring-degree. As described in \cite{CKL25}, the conversion to SinC encoding can be performed opportunistically during bootstrapping, by suitably adapting the bootstrapping circuit.

In coefficient encoding, ring-switching to the ring $\R_{N'}$, where $N'|N$, operation  decomposes a ciphertext $\ct = (a, b) \in \R_N^2$ to $k=N/N'$ ciphertexts $(a_j, b_j)_{0\le j < k}$ over the subring $\R_{N'}^2$. It proceeds by switching the secret key of $\ct$ to a secret $\sk'\in \R_{N'}$, obtaining a ciphertext $\ct' = (a', b')$ and decomposing the $a'$ and $b'$ parts as $a' = \sum_{j=0}^{k-1} a_j(X^k) X^j, $ $ b' =\sum_{j=0}^{k-1} b_j(X^k) X^j $. The resulting ciphertexts each encrypt $N'$ coefficients of the message to which $\ct$ decrypts. 

Let now $k'|N'$ be an integer and assume that the ciphertext $(a, b)$ encrypts a message $\mathbf{z} \in \mathbb{C}^{N/2}$ in $\SinC_{k',N}$ encoding; under the assumption  that $k' | k$, we show that the previous ring-switching process can also be used to handle $\SinC_{k',N}$-encoded ciphertexts. 
Let $\mathbf{z}^{(i)} = (z_{ik/2}, \dots, z_{(i+1)k/2-1})\in \mathbb{C}^{k/2}$. 
Following Eq.~\eqref{eq:sinc_ecd}, $(a, b)$ decrypts to $m \in \R_N$ with 
\begin{align*}
m & = \sum_{i=0}^{N/k'-1} \lfloor \Delta \cdot \textrm{iDFT}_{k'}(\mathbf{z}^{(i)}) \rceil X^i, \\
& = \sum_{j=0}^{k-1} \underbrace{\left(
\sum_{j' = 0}^{N'/k' - 1}
 \lfloor \Delta \cdot \textrm{iDFT}_{k'}(\mathbf{z}^{(kj'+j)}) \rceil (X^{k})^{j'} \right)}_{m_j} X^j.
\end{align*}

We notice that 
\[
m_j= \Ecd_{\SinC_{k',N'}}(\mathbf{z}^{(j)}, \mathbf{z}^{(j+k)}, \dots, \mathbf{z}^{(j+N/k'-k)})\in \R_{k'}^{N'/k'}\enspace.
\]

By switching the ring of $(a, b)$ to $\R_{N'}$ as previously, we obtain ciphertexts $(a_i, b_i)_{0\le i < k}$ in $\SinC_{k',N'}$ encoding. 

We use this mechanism to achieve the SinC format required to perform efficient CCMM. Specifically, we start with 8 ciphertexts in $\R_{2^{16}}$ each containing 32 matrices in $\mathbb{R}^{128 \times 128}$. These ciphertexts are moved through part of SlotToCoeff to obtain two ciphertexts in the $\SinC_{2^7, 2^{16}}$ encoding, and ring-switched  to $\R_{2^{12}}$. This gives 256 ciphertexts in $\SinC_{2^7, 2^{12}}$ encoding, each holding a fixed row of 16 different matrices. This is the format assumed by the batch-CCMM algorithm from~\cite{CKL25} for performing the 32 matrix multiplications in parallel.

\subsection{Slim polynomial evaluation for SoftMax}
\label{se:slim-poly}
We introduce an algorithm that fully utilizes the SIMD property of CKKS to evaluate a polynomial of degree $d$ on a ciphertext using only $\le N/(2d)$ slots with $O(\log d)$ key-switchings. 
An algorithm with the same goal was proposed in~\cite{OPP23}. 
This algorithm was designed for exact FHE schemes~\cite{b,fv,bgv}, and it leads to numerical instability for the CKKS scheme.

\smallskip
\noindent
{\bf Polynomial decomposition.}
Our algorithm relies on the following lemma. 
\begin{lemma}\label{le:sos}
  Let $P\in \RR[X]$ be a polynomial of even degree $d$ with positive leading
  coefficient. There exist two polynomials~$U$ and~$V$ of degree $\leq d/2$ and
  a scalar~$m$ such that $P(x) = U(x)^2 + V(x)^2 - m$. 
\end{lemma}
\begin{proof}
  We set $m = \max_{x\in \RR} (-P(x))$, which is finite under the assumption that~$P$ has a positive leading coefficient.
  Then~$P+m$ is a polynomial taking only non-negative values. As a result, there exist two polynomials $U$ and~$V$ such that $P+m= U^2 + V^2$ (see, e.g.,~\cite{Be17}).
\end{proof}

Contrary to the factorization $P = P_1\cdot P_2$, this decomposition guarantees that~$U$ and~$V$ cannot take very large values: for~$x$ in an interval ${\mathcal I}$, the values $|U(x)|$ and~$|V(x)|$ are bounded from above by $\sqrt{\max_{x\in {\mathcal I}} |P(x)| + m}$. 

We observe that two such polynomials can be computed from the roots of~$P+m$ using an efficient algorithm described, e.g., in~\cite{Be17}. Write
$P+m = p_0 \prod_{i=0}^{d-1} (X - \alpha_i)$; as $P+m$ is a positive polynomial
we can order the $\alpha_i$'s by pairs such that 
$\alpha_{2i} = \overline{\alpha_{2i+1}}$ (note that for our choice of $m$ in the proof of Lemma~\ref{le:sos}, the root~$\alpha_{2i}$ will be real
for some~$i$, but it then has even multiplicity).
By using the identity
\[
(X - \alpha_{2i}) (X - \alpha_{2i+1}) = (X - \textrm{Re}(\alpha_{2i}))^2 + \textrm{Im}(\alpha_{2i})^2 \enspace,
\]
we obtain $P$ as a product of sums of two squares. It then suffices
to  iteratively use the identity
\begin{align}
(U_0^2 + V_0^2)(U_1^2 + V_1^2) & = \left(\frac{U_0 U_1 - V_0 V_1 + U_0 V_1 + U_1 V
_0}{\sqrt{2}}\right)^2  \nonumber \\&+ \left(\frac{U_0 U_1 - V_0 V_1 - U_0 V_1 - U_1 V_0}{\sqrt{2}}\right)^2 \label{eq:sos}
\end{align}
to deduce the decomposition.

\smallskip 
\noindent
{\bf Polynomial evaluation.}
This gives the following encrypted algorithm for the evaluation of~$P$: starting with a slim ciphertext~$\ct$ encrypting $(m_i)_{0\le i < N/2}$ such that $m_{i+N/4} = m_i$, we evaluate~$U$ and~$V$ in parallel in slots of indices $0, \dots, N/4-1$ and $N/4, \dots, N/2-1$ to obtain $\ct'$, and deduce the value of~$P$ by computing $\ct'\odot \ct' + \Rot_{N/4}(\ct'\odot \ct')$. 
If the degrees of~$U$ and~$V$ are even, we can use the idea recursively: 
for a polynomial $P$ of degree $2^k$, we obtain a binary
tree of polynomials. We define $U_0^{(0)}$ to be $P$. Let us fix a parameter $j$ such that $0\le j < k$, and assume that we stop the recursion after $j$ steps. For $0\le \ell \le j$, the $\ell$-th level of the binary tree contains polynomials $U_0^{(\ell)}, \dots, U_{2^\ell-1}^{(\ell)}$ of degree $2^{k-\ell}$ and constants $m_0^{(\ell-1)}, \dots, m_{2^{\ell-1}-1}^{(\ell-1)}$ such that 
\begin{equation}\label{eq:recsos}
{U_{i}^{(\ell)}}^2 + {U_{i+2^{\ell-1}}^{(\ell)}}^2 = U_{i}^{(\ell-1)} + m_i^{(\ell-1)}, \  0\le i< 2^{\ell-1} \enspace.
\end{equation}

For $0\le \delta\le 2^{k-j}$, and given the number of slots $2^t$ of the target ciphertext such that $2^{t+j} \le N$, we define the plaintext $\v^{(\delta)}$ as containing in its slots $i2^t, ..., (i+1)2^t-1$ the degree-$\delta$ coefficient of the polynomial $U_{i\bmod 2^j}^{(j)}$. For $1\le \ell \le j$, we define the plaintext $\m^{(\ell-1)}$ as containing in its slots $i2^t, ..., (i+1)2^t - 1$ the value $m_{i \bmod 2^{\ell-1}}^{(\ell-1)}$.

If we cut the tree at level $j$, we obtain Algorithm~\ref{alg:slim}. 
This algorithm uses as a subroutine a function \textsf{Evaluate} which, on input a ciphertext~$\ct$ encrypting $(m_i)_{0\le i < N}$ and plaintexts $\v^{(0)}, \dots, \v^{(2^{k-j})}$ where $\v^{(t)}$ encodes a vector $(v^{(t)}_i)_{0\le i < N}$, returns $\ct'$ decrypting to $(\sum_{s=0}^{2^{k-j}} v^{(s)}_i m_i^s)_{0\le i < N}$. This function can be implemented using the Paterson-Stockmeyer algorithm, using plaintext rather than scalars as polynomial coefficients. Then, it uses $k+1$ multiplicative levels, and its cost is dominated by $O(2^{(k-j)/2}) + j$ ciphertext-ciphertext multiplications and $j$ rotations.
We refer to Appendix~\ref{app:slim-poly} for a more detailed analysis on the cost and numerical stability. 

\begin{algorithm}
\caption{\label{alg:slim} Slim polynomial evaluation algorithm for a polynomial of degree $2^k$, with depth $j < k$ recursion, on input a slim ciphertext with $2^t$ slots with $2^{t+j} \le N$.}
    \begin{algorithmic}[1]
\REQUIRE {$\ct = (a, b) \in {\R^2_{Q_{\ell}}}$}
\REQUIRE {Plaintexts $\v^{(0)}, \dots, \v^{(2^{k-j})}, \m^{(0)}, \dots, m^{(j-1)}$}
\REQUIRE {Key-switching keys $\rlk$, $(\rk_{2^{t+i}})_{0\le i < j}$}
\ENSURE{$(\ba', \bb')$.} 
    \STATE $\ct \leftarrow \mathsf{Evaluate}(\ct, \v^{(0)}, \dots, \v^{(2^{k-j})})$;
    \FOR{$\ell$ from $k-j$ downto $1$}
    \STATE $\ct \leftarrow \Mult_{\rlk}(\ct, \ct)$
    \STATE $\ct \leftarrow \ct + \Rot_{\rk_{2^{t+\ell-1}}}(\ct, 2^{t+\ell-1}) - \m^{(\ell-1)}$.
    \ENDFOR
    \RETURN $\ct$
    \end{algorithmic}
\end{algorithm}

\subsection{End-to-End Performance}
\label{se:cc-quick-result}

Table~\ref{tab:cc-quick-result} summarizes end-to-end prefill and decode latencies across GPU configurations. In the fully-encrypted 128-token setting, Sylph completes prefill in 20\,s on eight RTX PRO 6000 GPUs, achieving a $6.7\times$ speedup over Cerium on hardware that is approximately $5\times$ less costly. Against MOAI on the same hardware, the improvement exceeds $100\times$.\footnote{MOAI timing is estimated, as its memory requirement exceeds the capacity of a single RTX PRO 6000 GPU.} Sylph further supports a 3968 public + 128 private token setting, for which no prior homomorphic inference system reports results.

These results demonstrate that the combination of unified operator execution, outlier-aware calibration, and optimized homomorphic linear algebra substantially reduces the cost of fully encrypted Transformer inference.
 \section{Scaling to Long Heterogeneous Prompts}
\label{se:large-input}

\begin{figure*}
    \centering
    \includegraphics[width=2.1\columnwidth]{Figures/ctxt-algorithm-v260612.png}
    \caption{Overview of encrypted computation. The red circles and orange triangles indicate the bootstrap placements. The bootstrapping during SoftMax evaluation (orange triangles) is performed only in the narrow auxiliary track (see Section~\ref{se:shallow_Softmax}).}
    \label{fig:ct-alg}
\end{figure*}
This section extends the encrypted inference framework to long-context workloads using a heterogeneous prefill strategy. 
A major obstacle to long-context FHE inference is the cost of the prefill stage. 
The cost of prefill scales quadratically with the length of the input context. As a result, processing thousands of encrypted input tokens becomes prohibitively expensive.

However, many practical workloads contain a large amount of public context and only a small amount of private user input. Examples of public context include retrieved documents, shared conversation history, system prompts, or enterprise knowledge bases. In contrast, the user-specific query and instructions often constitute a much smaller but privacy-sensitive portion of the input. In what follows, we use the word "heterogeneous" to describe settings  that
combine public and private inputs and  the methods designed to address them. 

In this section, we propose a new framework to address workloads with such heterogeneous inputs. It covers heterogeneous prefill and decoding stages, and introduces new FHE algorithms for this setting. 

\subsection{Heterogeneous Public/Private Prefill}
We introduce a heterogeneous prefill mechanism that separates public and private prompts. A large non-sensitive context is processed in the clear, while the sensitive query and all query-dependent states remain protected.

For simplicity, we first describe the algorithm without FHE. 
We split the prompt of length $\ntok$ into two components: $P=(p_1,\ldots,p_{\ptok})$, the first public $\ptok$ tokens, and $S=(s_1,\ldots,s_{\etok})$, subsequent encrypted private $\etok$ tokens. Then, our plaintext algorithm is as follows. 
\begin{itemize}
    \item \textbf{Public prefill stage}: process the public prompt $P$ in the clear and generate a public KV cache $(K_{\mathrm{pub}},V_{\mathrm{pub}})$. It follows the standard Llama prefill. 
\item \textbf{Private prefill stage}: process the encrypted prompt $S$ and the public KV cache $(K_{\mathrm{pub}},V_{\mathrm{pub}})$ and generate the private KV cache $(K_{\mathrm{priv}},V_{\mathrm{priv}})$ and the first output token.  
    \item \textbf{Decode stages}: iteratively update the public and private KV caches and generate tokens; they are exactly the same as the standard Llama decode stages. 
\end{itemize}
The public and private prefill stages follow the standard Llama prefill stages, but with different input lengths. The standard prefill algorithm uses $(\ptok+\etok)$ tokens at once, and our prefill stages use $\ptok$ and $\etok$ tokens, respectively. 
Figure~\ref{fig:ct-alg} visualizes the proposed algorithm.

We now describe its homomorphic evaluation. The public prefill stage is performed in the clear, allowing us to fully leverage existing LLM implementations. 
Hence, we focus on the private prefill and decode stages. A key observation is that the additional computation induced by the public context is concentrated in the attention path (See the Attention Operation in Figure \ref{fig:ct-alg}). Other Transformer operators, (QKV-Projection, RoPE, O-Projection and FFN), operate on each token individually and are independent from the public KV cache; thus, algorithms discussed in Section~\ref{se:outlier} can be directly used for those operators.

The attention operation, on the other hand, mixes information \emph{across tokens}. Thus, the heterogeneous token types modify the entire attention computation. Specifically, the attention operation consists of three steps: QK step, SoftMax step, and ScoreV step. In the heterogeneous setting, the QK step computes both $ K_{\mathrm{pub}}\cdot Q^{T}$ and $K_{\mathrm{priv}}\cdot Q^{T}$, whose outputs are combined and passed to a SoftMax over $\ptok+\etok$ positions. The resulting attention scores are then applied to both $V_{\mathrm{pub}}$ and $V_{\mathrm{priv}}$ to produce the final attention output.
Importantly, this implies that the attention operation is evaluated over each of the public and private KV caches independently. 

Accordingly, we divide the private prefill into two components. The first component involves interactions between the encrypted queries and the public KV cache, and we denote it as \emph{plaintext-ciphertext attention (PC-attention)}. The second component involves the encrypted queries and the private KV cache, which we refer to as \emph{ciphertext-ciphertext attention (CC-attention)}. Figure~\ref{fig:ct-alg} illustrates an overview of the computation flow in our framework. 

\smallskip
\noindent \textbf{FHE design: private prefill stage.}
The different computational characteristics of PC-attention and CC-attention require distinct homomorphic computation strategies. 
\emph{CC-attention} operates exclusively on the encrypted private tokens, meaning that the algorithms and optimization techniques discussed in Section~\ref{se:outlier} can be directly adopted. 

In contrast, \emph{PC-attention} processes a massive, wide public KV cache. Importantly, while its input and output consist of a relatively small number of ciphertexts, the operation processes a massive\footnote{Under our specific parameters and implementation setup in Section~\ref{se:implem}, the input and output of PC-attention fit into 8 ciphertexts, while the PC-attention processes 256 ciphertexts during the computation.} number of intermediate ciphertexts during computation.
Performing bootstrapping over such a wide context would incur prohibitive overhead, creating a severe performance bottleneck.
Therefore, the core objective is to avoid bootstrapping as much as possible, ideally computing the PC-attention stage nearly bootstrapping-free. 

In Sections~\ref{se:pcmm}~and~\ref{se:softmax-changes}, we describe our specific designs for PCMM and SoftMax steps tailored for PC-attention. Together with the algorithms discussed in Section~\ref{se:outlier}, these components complete our entire long-context LLM inference pipeline. 

\smallskip
\noindent \textbf{FHE design: decoding stage.}
The decoding stage has the same structure as the prefill stage, and decomposes similarly in a PC-track and a CC-track. 
However, during the attention phase, matrix-matrix multiplications are replaced with matrix-vector multiplications, making the computational flow of uniform width. The circuit design is thus no longer constrained by bootstrapping overhead, and we systematically adopt the fastest algorithms for each component. In particular, we use~\cite{rhombus} for all PCMv steps, including during PC-attention. 

For CC-attention, we simply adopt same computation flow in Section~\ref{se:cc-quick-result}, as in the private prefill stage.

\subsection{PCMM for PC-attention}
\label{se:pcmm}

Given our goal of achieving a almost bootstrapping-free PC-attention design, a first challenge lies in executing massive linear algebra operations without invocating bootstrapping. 
To this end, we propose a PCMM variant of~\cite{JKLS18} that eliminates layout transitions and minimizes multiplicative depth.

Let $A$ and $B$ be matrices in $\mathbb{R}^{d\times d}$ and $C$ be their product. 
The JKLS matrix multiplication algorithm relies on the following (plaintext) matrix equation: 
\begin{equation}
\label{eq:JKLS}
C~=~\sum_{k=0}^{d-1} \rotcol^k( \sigma(A) ) \odot \rotrow^k( \tau(B) ) \enspace,
\end{equation}
where $\odot$ denotes the Hadamard (point-wise) multiplication between matrices stored as vectors by copying their rows one after another.  $\rotrow$ and $\rotcol$ respectively represent rotation of row and column, 
and~$\sigma$ and~$\tau$ represent relevant permutations. Specifically, we have, for all~$0 \leq i,j <d$:
\begin{align*}
\left(\rotcol^k(A)\right)_{i,j} = A_{i,j+k\bmod{d}} \enspace,\enspace& 
\left(\sigma(A)\right)_{i,j} = A_{i, i+j\bmod{d}}\enspace,\\
\left(\rotrow^k(A)\right)_{i,j} = A_{i+k\bmod{d},j}\enspace,\enspace
&\left(\tau(A)\right)_{i,j} = A_{i+j\bmod{d},j}\enspace.
\end{align*}

The original JKLS method homomorphically evaluates Equation~\eqref{eq:JKLS} to obtain an encryption of~$C$ from encryptions of~$A$ and~$B$. It consumes $3$ multiplicative levels: one for $\sigma$ and $\tau$ (performed in parallel), one for the column rotations (row rotations are implemented with a regular CKKS rotation and do not consume a level, if we write the matrix row after row in slots), and one for Hadamard multiplications.

\smallskip
\noindent\textbf{Low-depth, low-memory PCMM}. 
While $\PCMM$ can be viewed as a special case of $\CCMM$ where the naive JKLS algorithm consumes three levels, we reduce this multiplicative depth to one by exploiting $\PCMM$-specific properties. First, by setting~$A$ as the plaintext weight matrix and~$B$ as the ciphertext activation matrix, we eliminate one multiplicative level, as plaintext column rotations are depth-free. 

Second, we adaptively update the packing structure of matrices to further reduce the depth. Observe that the naive JKLS algorithm consumes only a single level if the input matrix~$B$ is formatted as $\tau(\ct_B)$ instead of $\ct_B$. Based on this observation, we modify the JKLS algorithm to take $\tau^{\ell+1}(\ct_B)$ as input and output $\tau^{\ell}(\ct_C)$ for an arbitrary integer~$\ell$.

We now describe our depth-optimized PCMM with BSGS optimization. Let $\pt_A$ and $\tau^{\ell+1}(\ct_B)$ be a plaintext and a ciphertext that respectively store the input matrices~$A$ and~$\tau^{\ell+1}(B)$.
For arbitrary integers~$g,b$ with~$g\cdot b =d$, we compute $\tau^{\ell}(\ct_C)$ as 
\begin{equation}\label{eq:pcmm}
 \sum_{j=0}^{g-1} \rotrow^{j\cdot b} \left( \sum_{i=0}^{b-1} \pt_{A,i,j,\ell} \odot \rotrow^i(\tau^{\ell+1}(\ct_{B}))\right) \enspace,
\end{equation}
where the plaintext matrix $\pt_{A,i,j,\ell}$ is a rearranged version of~$\pt_A$ defined as  
$\pt_{A,i,j,\ell}= \rotrow^{-\ell\cdot(i+j\cdot b)-j\cdot b} \circ \rotcol^{i+j\cdot b} (\tau^{\ell}\circ\sigma(\pt_{A}))$. 
We refer to Appendix~\ref{app:pcmm} for the correctness of this equation. 

Eq.~\eqref{eq:pcmm} yields an algorithm that computes $\tau^{\ell}(\ct_C)$ from~$\pt_A$ and $\tau^{\ell+1}(\ct_B)$ using $1$ multiplicative level and $\mathcal{O}(\sqrt{d})$ ciphertext rotations. 

Equation~\eqref{eq:pcmm} requires multiple rearranged plaintexts $\pt_{A,i,j,\ell}$. However, storing all precomputed plaintexts is impractical for large-scale models due to the massive memory overhead.  To tackle this issue, we store only $\tau^\ell\circ\sigma(\pt_A)$ and compute $\pt_{A,i,j,\ell}$ for each $i$,$j$ and~$\ell$ at runtime. This approach requires additional plaintext computations, but it significantly reduces the memory requirements. 
This is applicable to other $\PCMM$ (and PCMv) algorithms such as Powerformer~\cite{PLL25} with a BSGS structure. 
However, our proposed $\PCMM$ algorithm requires fewer plaintext operations since each $\pt_{A,i,j,\ell}$ can be obtained with only two plaintext rotations from $\tau^\ell\circ\sigma(\pt_A)$ compared to 8 rotations required by \cite{PLL25} .

\medskip\noindent \textbf{Application to PC-attention}.
PC-attention consists of a $\PCMM_1$ (QK step), a Softmax step, and $\PCMM_2$ (ScoreV step).
Both PCMM steps are performed using the previous algorithm, in slot-encoding, thus avoiding SlotToCoeff, CoeffToSlot, and bootstrapping  during the wide PC-attention. 

As each $\PCMM$ call simultaneously evaluates $\tau^{-1}$, 
we apply $\tau^2(\cdot)$ right after RoPE, before the computation becomes wide. This allows us to obtain the correct layout at the end of the attention phase. 
For the remaining packing layouts, we use the same strategy as in Section~\ref{se:pipeline}.

As SoftMax is evaluated between the two $\PCMM$ steps, we perform it on a matrix $M$ represented as~$\tau(M)$:
$$\left(\mathsf{SoftMax}((\tau(M)_{j,0})_j),\cdots,\mathsf{SoftMax}((\tau(M)_{j,d-1})_j)\right)\enspace.$$
Each vector $(\tau(M)_{j,i})_j$ is the $i$-th column of $M$ rotated by $i$: 
$$\left(M_{i,i},\cdots,M_{d-1,i},M_{0,i},\cdots,M_{i-1,i}\right)\enspace,$$ 
and the value $\mathsf{SoftMax}((\tau(M)_{j,i})_j)$ is $\mathsf{SoftMax}((M_{j,i})_j)$ rotated by $i$, so that the output is
$$\tau\left(\mathsf{SoftMax}((M_{j,0})_j),\cdots,\mathsf{SoftMax}((M_{j,d-1})_j)\right)\enspace.$$
allowing the next $\PCMM$ to proceed on SoftMax output.

\subsection{How SoftMax Changes}
\label{se:softmax-changes}
\medskip\noindent \textbf{Increasing the dimension of SoftMax instances}
The heterogeneous prompt increases the dimension of each encrypted SoftMax instance from $d = \etok$ to $d = \ptok + \etok$, while the number of encrypted SoftMax instances to be handled in parallel remains equal to $\ell = \etok$.
In the context of the SoftMax algorithm from~\cite{CHKPS24} (see Section~\ref{se:shallow_Softmax}), the cheap and shallow main track thus operates on $\etok~(\ptok + \etok)$ inputs while the deep and expensive auxiliary track operates only on $\etok$ inputs. The large public/private prompt, while increasing significantly the size of the input to SoftMax, has thus limited impact on SoftMax costs. In our experiments ($\ptok = 31\etok$), a $32\times$ increase on SoftMax input size leads only to a $3\times$ increase in total computation time. 

\medskip\noindent \textbf{Designing a shallow SoftMax}
To minimize bootstrappings in the main track (which, for large input prompt, manipulates many ciphertexts during the PC-attention phase) we need to keep the number of iterations of the normalize-and-square strategy (see~Section~\ref{se:shallow_Softmax}) small. 
Rather than relying on worst-case bounds as~\cite{CHKPS24}, we use the distributional data computed during calibration (see Section~\ref{sec:model_calibration}) to obtain sharp estimates on the range of the inverse square root computations. These estimates allow us to restrict to two iterations, yielding a SoftMax implementation using only 8 levels in its main track.

 \section{Sylph System Implementation}\label{se:implem}
This section focuses on the multi-GPU aspects of the implementation of Sylph, a multi-GPU runtime for encrypted Transformer inference.
Sylph realizes the homomorphic operators introduced in the previous sections by distributing computation across GPUs in a way that reduces memory pressure and inter-GPU traffic.

\subsection{Data Partitioning}\label{se:distrib}
We parallelize PCMM across multiple GPUs using a two-dimensional partitioning over the input and output dimensions. Our main goal is to distribute the plaintext weight matrices as much as possible: they need to reside in GPU memory throughout the computation and represent a large proportion of the memory footprint.

Let $W = (w_{i,j})_{0\le i < d, 0\le j < d'} \in \mathbb{R}^{d\times d'}$ be a plaintext weight matrix used during a PCMM step of the form $B = W\cdot A$, where $A = (a_{ij})_{0\le i < d', 0\le j < d''}$ and $B = (b_{ij})_{0\le i < d, 0\le j < d''}$. 

Given $g$ GPUs, we let $p_r$ and $p_c$ be two positive integers such that $p_r|d$, $p_c|d'$ and $p_r \cdot p_c = g$, and split $W$ blockwise as 
\[
W^{(u, v)} = (w_{i+ud/p_r, j+vd'/p_c})_{0\le i < d/p_r, 0\le j < d'/p_c}, 
\]
and accordingly
\begin{align*}
A^{(v)} & = (a_{i+vd'/p_c,j})_{0\le i < d'/p_c, 0\le j < d''},\\ 
B^{(u)} & = (b_{i+ud/p_r, j})_{0\le i < d/p_r, 0\le j < d''}, 
\end{align*}
for $0\le u < p_r$, $0\le v < p_c$. 

Each GPU is assigned a pair $(u, v)$ as a logical coordinate. The GPU numbered $(u, v)$ stores the plaintext block $W^{(u,v)}$ and the activation values $A^{(v)}$, and computes the term $W^{(u, v)}\cdot A^{(v)}$. In view of
\begin{equation}\label{eq:blockmat}
B^{(u)}  = \sum_{0\le v < p_c} W^{(u,v)}\cdot A^{(v)}, 
\end{equation}
the PCMM result is obtained by gathering, for all $u$, the products $W^{(u, v)} \cdot A^{(v)}$ across the nodes $(u, v)$. 

This partitioning distributes
PCMM computation across all GPUs, with each GPU stores approximately $1/g$ of the weight matrix, and a proportion $1/p_c$ of the activation input. 
This partitioning describes the logical matrix-level view; the actual ciphertext representation follows the packing layouts defined in Section~\ref{se:pipeline}. 
For PCMV, we use the one-dimensional special case with $p_c=1$, where the activation is not partitioned and only the plaintext matrix is distributed across GPUs.

\subsection{Inter-GPU communication}\label{se:comm}
Our design requires communication to accumulate the block products, as shown by Eq.~(\ref{eq:blockmat}). For $0\le v' < p_c$, GPUs $(u, v')$ share and accumulate the block products $W^{(u, v')}\cdot A^{(v')}$ through a reduce-scatter operation: each of the $p_c$ GPUs sends  to share a proportion $1/p_r$ of the total result. 

In our homomorphic computational flow, each block product is represented as degree-$2^{12}$ ciphertexts. After reduce-scatter, each GPU retains the complete ring-degree-$2^{12}$ ciphertexts that it owns, composes them into ring-degree-$2^{16}$ ciphertexts, and bootstraps them locally.

The placement of linear algebra steps is such that the block products shared during the communication stage are at the bottom level, leading to low communication.

\subsection{Model Conversion}
We convert the Llama-3-8B weights into the
layouts directly consumed by \cite{BCHPS24} and \cite{rhombus}.  The conversion is primarily a layout transformation: weights are partitioned and tiled. \cite{rhombus} consumes coefficient-domain plaintext polynomials, hence further requires transforming the plaintext polynomials into the NTT domain. 

Table~\ref{tab:model-conversion-memory} summarizes the resulting memory
footprint for the converted Llama-3-8B projection weights across all transformer layers. As the original 16-bit weights are stored under a $\approx 64$-bit modulus, we observe a $\approx \times 4$ storage expansion overall.

\begin{table}[t]
\centering
\caption{Memory footprint of converted Llama-3-8B projection weights. The
baseline stores weights in 16-bit precision.}
\label{tab:model-conversion-memory}
\begin{tabular}{lccc}
\hline
Representation & Total memory & Expansion \\
\hline
Original 16-bit weights  & $13.0$ GiB & $1.00\times$ \\
\cite{BCHPS24} packed PCMM tiles  & $52.0$ GiB & $4.00\times$ \\
\cite{rhombus} NTT plaintexts & $58.0$ GiB & $4.49\times$ \\
\hline
\end{tabular}
\end{table}

\subsection{Experimental Setup}
RTX 6000 experiments are conducted on a server with eight NVIDIA RTX PRO 6000 GPUs and two Intel Xeon 6530P (PCIe Gen5, 512GB memory). B200 experiments are carried on a server equipped with four B200 GPUs interconnected through NV18 NVLink and two AMD EPYC 9365 CPUs (NVLink NV18, 2.2TB memory). Both systems use the HEaaN2 FHE library~\cite{heaan2} for CKKS primitive operations and cryptographic setup. 

\subsection{Experimental Results}

Table~\ref{tab:pipeline-time} reports the per-layer execution time of Sylph on Llama3 for prefill ($T=128$) and decode ($T=1$). The dominant costs come from PCMM and CCMM kernels. With 8 GPUs, prefill QKV projection decreases from 326\,ms to 65\,ms, O projection from 315\,ms to 40\,ms, and FFN gate/up projection from 1040\,ms to 132\,ms, showing effective scaling of large PCMMs. This (almost) linear speedup illustrates the efficiency of our distribution and communication strategies~(Secs.~\ref{se:distrib} and~\ref{se:comm}).

The encrypted attention kernel also benefits from distribution. In prefill, QK, SoftMax, and ScoreV decrease from 316/440/452\,ms to 41/81/63\,ms, respectively. Compared with the public-score PC-operation baseline on 8 GPUs, the CC-operation path is much faster for QK and ScoreV, avoiding the high communication and materialization cost of public attention scores.

Overall, Sylph reduces per-layer prefill latency from 3.55\,s on one GPU to 0.55\,s on 8 GPUs, achieving a 6.5$\times$ speedup. Decode is less scalable because $T=1$ limits parallelism, but still improves from 0.98\,s to 0.56\,s per layer.

\begin{table}[t]
\centering
\scriptsize
\renewcommand{\arraystretch}{1.08}
\setlength{\tabcolsep}{2.5pt}
\caption{Per-layer GPU time for Sylph's homomorphic Llama3 evaluation. We use $T=128$ for prefill and $T=1$ for decode, $H=4096$, $I=14336$, $h=32$, $d=128$.}
\label{tab:pipeline-time}
\begin{tabular}{@{}llrrrr@{}}
\toprule
& & \multicolumn{2}{c}{\textbf{Prefill (ms)}} & \multicolumn{2}{c}{\textbf{Decode (ms)}} \\
\cmidrule(lr){3-4} \cmidrule(lr){5-6}
\textbf{Step} & \textbf{Shape} & \textbf{1 GPU}  & \textbf{8 GPU} & \textbf{1 GPU} & \textbf{8 GPU} \\
\midrule
\multicolumn{6}{@{}l}{\textbf{QKV projection}} \\
Pre-attn norm  & $T{\times}H$                              & 20 & 19 & 9 & 10\\
QKV proj.      & $(T{\times}H)(H{\times}6144)$             & 326  & 65 &135 & 70\\
RoPE           & $T{\times}5120$                           & 35 & 12   &  2  & 2\\
\midrule
\multicolumn{6}{@{}l}{\textbf{Attention CC-operation}} \\
QK scores      & $(T{\times}d)(d{\times}T){\times}h$       & 316  & 41&154& 124\\
SoftMax        & $T{\times}T{\times}h$                     & 440  & 81   & 49& 50\\
ScoreV         & $(T{\times}d)(d{\times}T){\times}h$       & 452 & 63   & 155& 111\\
\midrule
\multicolumn{6}{@{}l}{\textbf{Attention PC-operation}} \\
QK scores, pub.  & $(T{\times}d)(d{\times}T){\times}h$       & -  & 586 & - & 70\\
SoftMax          & $T{\times}4096{\times}h$                  & -  & 153  & - & 123\\
ScoreV, pub.     & $(T{\times}3968)(3968{\times}d){\times}h$ & -  & 635 & -& 83\\
\midrule
\multicolumn{6}{@{}l}{\textbf{O projection}} \\
O proj.          & $(T{\times}H)(H{\times}H)$                 & 315  & 40   & 80& 40\\
\midrule
\multicolumn{6}{@{}l}{\textbf{Feed-forward network}} \\
Pre-FFN norm     & $T{\times}H$                              &  20  & 20   & 20& 20\\
Gate/up proj.    & $(T{\times}H)(H{\times}2I)$               & 1040  & 132  & 133& 57\\
Gate act.        & $T{\times}I$                              & 198 & 27   & 4& 4\\
Down proj.       & $(T{\times}I)(I{\times}H)$                & 391   & 50   & 71& 45\\
\bottomrule
\end{tabular}\end{table} \section{Conclusion}
In this work, we scaled up FHE-based LLM inference towards meeting practical deployment demands. We systematically co-designed efficient FHE circuits for Llama-3-8B inference by orchestrating diverse FHE algorithms across Transformer operators. To this end, we developed tailored algorithms and incorporated machine learning techniques to carefully manage the dynamic range and precision of intermediate variables. Furthermore, to scale the supported input length, we accommodated heterogeneous user queries consisting of both private (encrypted) and public (cleartext) token segments. Finally, we instantiated these techniques in \emph{Sylph}, a GPU-accelerated implementation of our system. Experimental results demonstrate that Sylph significantly outperforms the state of the art on homogeneous, fully encrypted baseline tasks, while successfully scaling the total context length through our heterogeneous prompt processing.

We emphasize that this work primarily focused on the prefill stage, which represents the most computationally intensive stage of Llama-3-8B inference. While our decoding stage already outperforms existing baselines, its performance gains are more modest compared to those achieved in the prefill stage. We envision that various machine learning optimizations tailored for decoding, such as speculative decoding, can be integrated into our framework to further accelerate the decoding stage.

\bibliographystyle{IEEEtran}
\bibliography{bibliography}

\newpage 
\appendices

\section*{Acknowledgments}
The authors used Claude Code 2.1.173 to assist with the programming implementation. After using this tool, the authors thoroughly reviewed and take full responsibility for the final code.

This paper was edited for spelling and grammar using Gemini, ChatGPT, Claude, Grammarly, Overleaf's AI Assist. We used ChatGPT for generating skeletons of tables in LaTeX. 

\section*{Ethics considerations}
This work focuses strictly on the cryptographic optimization and performance evaluation of an open-weight large language model (Llama-3-8B) in a privacy-preserving context using Fully Homomorphic Encryption (FHE/CKKS). We think that our research does not introduce ethical concerns of risks of harm (no human subjects, no collection of personally identifiable information, no use of proprietary data). 

\section{CKKS functionalities}\label{app:CKKS}
\subsection{Encryption and decryption}
CKKS uses RLWE ciphertext formats~\cite{SSTX09,LPR10}\footnote{Some functionalities require moving to MLWE formats~\cite{BGV14,LS15} temporarily.} 
\begin{itemize}
\item[$\bullet$] \textsf{SetUp}: chooses a tuple $(N, P, Q, h, {\mathcal D})$, where $N, P, Q,$ and $h$ are positive integers (respectively ring degree, auxiliary modulus, top ciphertext modulus, and secret key Hamming weight) and ${\mathcal D}$ is a distribution on~$\R$. These parameters are chosen so that RLWE with ring degree~$N$ and modulus~$PQ$ with sparse ternary secrets of Hamming weight~$h$ and error distribution~${\mathcal D}$ has 128-bit security. \textsf{Setup} also picks and returns a uniform $\sk \in \ZZ[X]/(X^N+1)$ with coefficients in $\{-1, 0, 1\}$ and~$h$ non-zero coefficients. The choice of $(N, P, Q)$ among secure values takes into account the circuit to be implemented, in particular its shape in terms of width (related to~$N$),  depth (related to~$Q$), and number of key switchings (the cost of which is related to an integer parameter $\dnum \approx (\log Q)/(\log P)$). 
\item[$\bullet$] $\enc_{\sk, {\mathcal D}, Q}(\pt)$: on input $\pt \in \ZZ[X]/(X^N+1)$, return a pair $(a, -a \cdot \sk + \pt + e \bmod Q)$, where $e \leftarrow {\mathcal D}$ and $a$ is uniformly chosen in $\R_Q = \R/Q\R$; 
\item[$\bullet$] $\dec_{\sk}(\ct)$: on input $\ct = (a, b) \in \R_Q^2$, return $a\cdot \sk + b \bmod Q$.
\end{itemize}
In particular, the CKKS ciphertext space is~$\R_Q^2$. For simplicity, we omit the possibility to enable public key encryption, and encryption/decryption with various moduli. 
In the sequel, we shall simply write $\enc$ and~$\dec$, and assume that all their parameters are fixed at setup time. 

An important property of RLWE ciphertexts for homomorphic encryption is the ability to switch keys. A \emph{key-switching key} from the secret key $\sk$ to the secret key~$\sk'$ is an encryption of $P\cdot \sk'$ under~$\sk$ in~$\R_{PQ}^2$, namely a pair $(a, -a\cdot \sk + P \cdot \sk' + e)$, were $e\leftarrow {\mathcal D}$ and $a$ is uniformly chosen in~$\R_{PQ}$ (assuming that $\dnum=1$). This key-switching key, which can be made public, is used in the key-switching operation:
\begin{itemize}
    \item[$\bullet$] $\KS_{\ksk}(\ct)$: on input a key-switching key $\ksk$ from~$\sk$ to~$\sk'$ in $\R_{PQ}^2$ and a ciphertext $\ct \in \R_{Q}^2$ encrypted under~$\sk$, returns a ciphertext $\ct' \in \R_{Q}^2$ under~$\sk'$ such that
    $\dec_{\sk'}(\ct') \approx \dec_{\sk}(\ct)$. $\KS$ reduces to arithmetic in~$\R_{PQ}$, with a number of operations depending on~$\dnum$.
\end{itemize}

\subsection{Homomorphic functionalities}
CKKS offers the following homomorphic functionalities.
\begin{itemize}
    \item[$\bullet$]$\Add(\ct_1, \ct_2)$: on input two ciphertexts $\ct_1$ and $\ct_2$ in~$\R_Q^2$, return a ciphertext $\ct \in \R_Q^2$ such that $\dec(\ctres) \approx \dec(\ct_1) + \dec(\ct_2)$; this operation is implemented as componentwise addition in $\R_Q^2$; 
    \item[$\bullet$]$\PCMult(\pt, \ct)$: on input a plaintext $\pt \in \R$ and a ciphertext $\ct \in \R_Q$, both with scaling factor $\Delta$, return a ciphertext $\ct' \in \R_{Q'}^2$ with $Q' \approx Q/\Delta$, such that $\dcd(\dec(\ct')) \approx \dcd(\dec(\ct)) \odot \dcd(\pt)$.
    \item[$\bullet$] $\Mult_\rlk(\ct_1, \ct_2)$: on input two ciphertexts $\ct_1$  and $\ct_2$ in~$\R_Q^2$ with scaling factors $\Delta$, and a key-switching key $\rlk \in \R_{PQ}^2$ from $\sk^2$ to $\sk$, return a ciphertext $\ct' \in \R_{Q'}^2$ with $Q'\approx Q/\Delta$, such that $\dcd(\dec(\ct')) \approx \dcd(\dec(\ct_1)) \odot \dcd(\dec(\ct_2))$. 
    \item[$\bullet$]$\Rot_{\rk_k}(\ct, k)$: on input a ciphertext $\ct$, an integer $0 \leq k< N/2$, and a switching key from $\sk(X^{5^k})$ to $\sk$, return a ciphertext $\ct'$ such that $\dcd(\dec(\ct'))$ is a cyclic rotation of $\dcd(\dec(\ct))$ by $k$ positions.
\end{itemize}

\subsubsection{Rescaling, precision, and modulus handling}
\label{se:modulus}
The use of a scaled encoding with scaling factor $\Delta$ implies that the arithmetic result of an encoded multiplication must be corrected via division by $\Delta$ in order to restore the correct scaling factor. This is implemented homomorphically as $\RS$:
\begin{itemize}
\item[$\bullet$] $\RS(\ct)$: on input a ciphertext $\ct \in \R_Q^2$, returns $\ct' \in \R_{Q'}^2$ with $Q' \approx Q/\Delta$, such that $\dec(\dcd(\ct')) \approx \Delta^{-1}\cdot \dec(\dcd(\ct))$. 
\end{itemize}

As a consequence, every multiplication reduces the modulus of a ciphertext by a factor $\approx \Delta$. This is usually accounted via 
the notion of \emph{level}: given integers $Q_0, Q_1, \ldots, Q_L$ where $Q_i/Q_{i-1} \approx \Delta$ for $i \ge 1$, we define level-$j$ ciphertexts as elements of~$\R_{Q_j}^2$. Multiplication (both $\PCMult$ and $\Mult$) can then be seen as taking its inputs in~$\R_{Q_j}^2$ and returning its result in~$\R_{Q_{j-1}}^2$. 

The numerical model of CKKS is close to a fixed-point number system, meaning that the precision of an input value $x$ is absolute and does not depend on the order of magnitude of $x$. The precision of CKKS computations is driven by the scaling factor~$\Delta$.
 Choosing~$\Delta$ is a tradeoff between precision needs and modulus usage a larger~$\Delta$ provides better accuracy but exhausts ciphertext modulus faster. 

\section{Visualization of Llama Inference Pipeline}
\label{app:figure}
\begin{figure}[htbp]
    \centering
    \includegraphics[width=\linewidth]{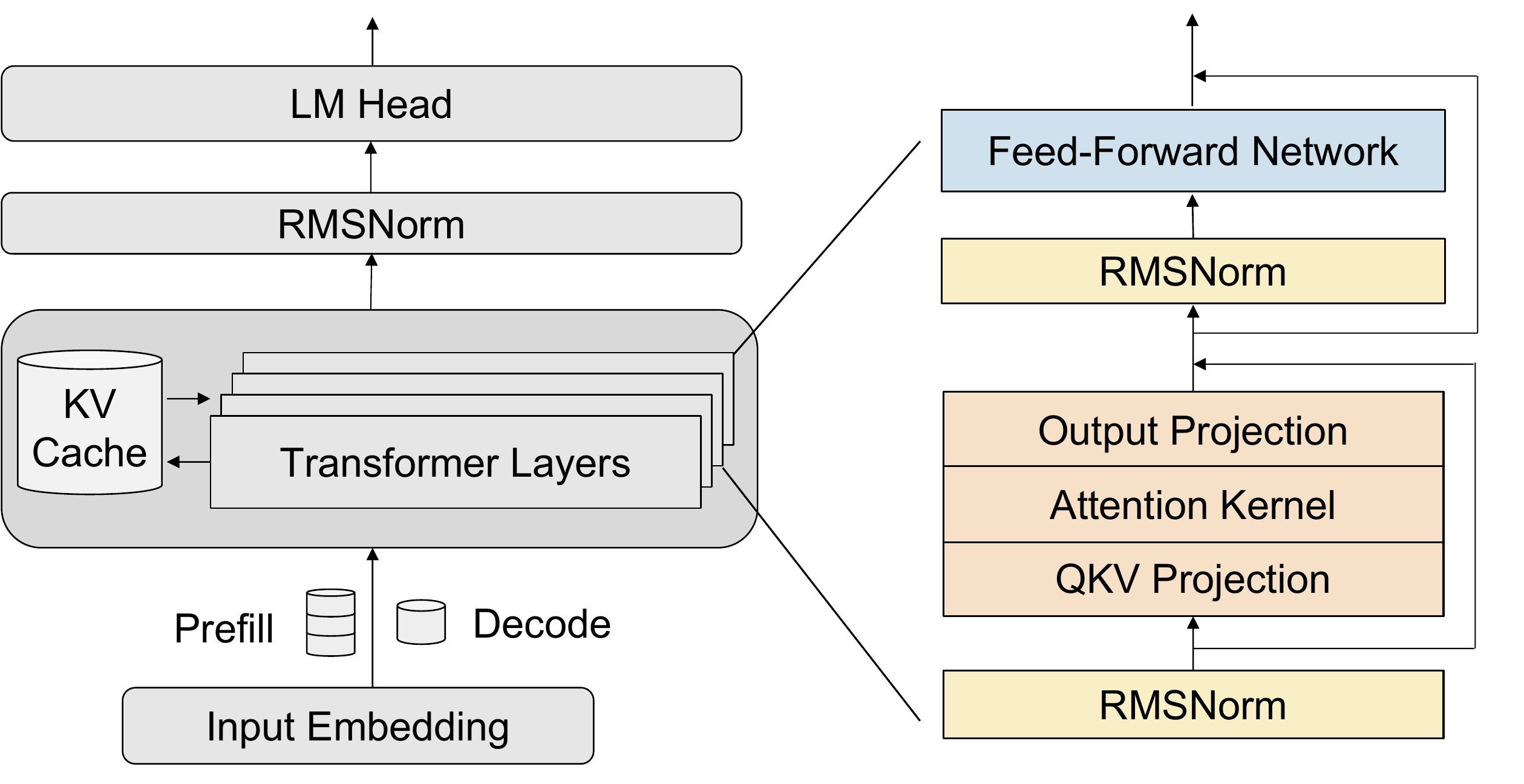}
    \caption{Llama inference pipeline with KV caching. The left figure shows the overall autoregressive inference flow, while the right figure illustrates the structure of a Transformer layer.}
\label{fig:pt-alg}
\end{figure}

Figure~\ref{fig:pt-alg} illustrates the Llama inference pipeline. 

\section{Perplexity under Various Precisions}
\label{app:perp}
We report the perplexity of Llama inference under various precision settings. 
\begin{table}[htbp] \centering \caption{Perplexity under different precision settings.} \label{tab:precision_perplexity} \begin{tabular}{lccccc} 
\toprule 
Precision & $2^{-9}$ & $2^{-10}$ & $2^{-11}$ & \textbf{$2^{-12}$} & FP16 \\ \midrule 
Perplexity & 6.33 & 6.15 & 6.14 & \textbf{6.13} & 6.13 \\ 
\bottomrule 
\end{tabular} 
\end{table} 

\section{Analysis of Algorithm~\ref{alg:slim}}
\label{app:slim-poly}
We analyze the correctness, computational cost, and numerical stability of  Algorithm~\ref{alg:slim}. 
\begin{theorem}\label{th:slim}
    Algorithm~\ref{alg:slim} is correct. It uses $k+1$ multiplicative levels, and its cost is dominated by $O(2^{(k-j)/2}) + j$ ciphertext-ciphertext multiplications and $j$ rotations.
\end{theorem}
\begin{proof}
Correctness follows from the discussion in Section~\ref{se:slim-poly}.  Using the Paterson-Stockmeyer algorithm, \textsf{Evaluate} requires $k-j+1$ computational levels and $O(2^{(k-j)/2})$ operations, while the following loop requires $j$ multiplicative levels, $j$ multiplications and $j$ rotations. 
\end{proof}

By obtaining on input a ciphertext containing $\left(v^{(2^{k-j})}_i\right)^{2^{j-k}} \cdot \ct_i$ by fusing a plaintext-ciphertext multiplication in a previous operation, Algorithm~\ref{alg:slim} can be made to use only $k$ levels.

\smallskip 
\noindent
{\bf Numerical stability.}
As $|x^2 - y^2| \le 2\max(|x|, |y|) \cdot |x - y|$, every iteration of the main loop of Algorithm~\ref{alg:slim} introduces a numerical error of 
$\approx 1 + \log(|U_{i}^{(\ell)} (x)| + |U_{i+2^{\ell-1}}^{(\ell)} (x)|)$ bits in slot~$i$.

As the proof of Lemma~\ref{le:sos} shows that we can obtain different decompositions by changing the  ordering of the roots of the polynomial $P$, it is thus preferable to search for a decomposition such that, at each level of the recursion, the polynomials~$U$ and~$V$ are small over the interval of approximation. In view of Equation~\eqref{eq:recsos}, exploring the set of possible decompositions at step $\ell-1$ in search of~$U_i^{(\ell-1)}$ with a small associated~$m_i^{(\ell-1)}$ is likely to yield polynomials~$U_i^{(\ell)}$ and~$U_{i+2^{\ell-1}}^{(\ell)}$ with smaller values at the next stage.

\section{Correctness of PCMM in Section~\ref{se:pcmm}}
\label{app:pcmm}
We start with a lemma:
\begin{lemma}
\label{le:tau}
Let $M \in \RR^{d\times d}$. For each integer~$k$, we have: 
    $$\tau\left(\rotrow^k(M)\right) = 
    \rotrow^k(\tau(M))\enspace,$$
    and
    $$\tau\left(\rotcol^k(M)\right) = \rotrow^{-k}\circ\rotcol^{k}(\tau(M))\enspace.$$ 
\end{lemma}
\begin{proof}
    Let $0 \leq i,j <d$. We have 
    \begin{align*}
    \left(\tau\left(\rotrow^k(M)\right)\right)_{i,j}
    &= \left(\rotrow^k(M)\right)_{i+j, j} = M_{i+j+k, j}\\&=\left(\tau(M)\right)_{i+k,j}= \left(\rotrow^k(\tau(M))\right)_{i,j}
    \end{align*}
    and

    \begin{align*}
    &\left(\tau\left(\rotcol^k(M)\right)\right)_{i,j}
    = \left(\rotcol^k(M)\right)_{i+j,j} = M_{i+j, j+k}\\&=\left(\tau(M)\right)_{i-k,j+k}=\hspace*{-.05cm}\left(\rotrow^{-k}\hspace*{-.05cm}\circ\rotcol^{k}(\tau(M))\right)_{i,j}\enspace,
    \end{align*}
where all index operations are modulo~$d$.
\end{proof}

Using Lemma~\ref{le:tau}, we proceed with Theorem~\ref{thm:pcmm}, which modifies the original JKLS algorithm by applying $\tau^{\ell}$, to homomorphically compute $\tau^{\ell}(C)$ from $\tau^{\ell+1}(B)$. 

\begin{theorem}
\label{thm:pcmm}
Let $A,B \in \RR^{d\times d}$ and $C = A \cdot B$. For each integer $\ell$, we have
$$
\tau^{\ell}(C)=\sum_{k=0}^{d-1} \left(\rotrow^{-\ell\cdot k}\circ\rotcol^{k}( \tau^{\ell}\circ\sigma(A) ) \right) \odot \left(\rotrow^k( \tau^{\ell+1}(B) )\right) \enspace.
$$
\end{theorem}
\begin{proof}
We apply $\tau^{\ell}$ to both sides of Equation~\eqref{eq:JKLS}: 
\begin{align*}
\tau^{\ell}(C)&=\tau^{\ell}\left(\sum_{k=0}^{d-1} \rotcol^{k}(\sigma(A)) \odot \rotrow^k( \tau(B) )\right) \\
&=\sum_{k=0}^{d-1} \tau^{\ell}\left(\rotcol^{k}(\sigma(A))\right) \odot \tau^{\ell}\left(\rotrow^k( \tau(B) )\right)\enspace.
\end{align*}
Since the $\rotrow$ and~$\tau$ operators commute by Lemma~\ref{le:tau}, we have:
\begin{align*}
\tau^{\ell}\left (\rotrow^k( \tau(B) )\right)
=\rotrow^k\left( \tau^{\ell+1}(B) \right)\enspace.
\end{align*}
Moreover, Lemma~\ref{le:tau} implies that
\begin{align*}
\tau^{\ell}\left(\rotcol^{k}(\sigma(A))\right) 
&= \tau^{\ell-1}\left(\tau\left(\rotcol^{k}(\sigma(A))\right)\right)
\\&= \tau^{\ell-1}\left(\rotrow^{-k}\circ
\rotcol^{k}( \tau\circ\sigma(A) ) \right)
\\&= \rotrow^{-k}\left(\tau^{\ell-1}\left(
\rotcol^{k}( \tau\circ\sigma(A) ) \right)\right)
\\&= \rotrow^{-2k}\left(\tau^{\ell-2}\left(
\rotcol^{k}( \tau^2\circ\sigma(A) ) \right)\right)
\\&=\cdots= \rotrow^{-\ell \cdot k}\circ
\rotcol^{k}\left( \tau^\ell\circ\sigma(A) \right)\enspace.
\end{align*}
This completes the proof.
\end{proof}

\end{document}